\documentclass[twocolumn,showpacs,superscriptaddress,preprintnumbers,amsmath,amssymb,longbibliography,floatfix,aps]{revtex4-1}
\usepackage{graphicx}
\usepackage{dcolumn}
\usepackage{bm}
\usepackage{amsmath}
\usepackage{color}
\usepackage{wrapfig}
\usepackage{lipsum}

\newcommand{\hatp}{\hat{P}}
\newcommand{\hatw}{\hat{\omega}}
\newcommand{\hatg}{\hat{\gamma}}
\newcommand{\hata}{\hat{\alpha}}
\newcommand{\wc}{\hat{\omega}_c}
\newcommand{\hatc}{\hat{c}}

\begin{document}

\title{Proximity to Jamming Governs Acoustic Attenuation in Damped Packings}

\author{Colton Kawamura}
\affiliation{Van der Waals-Zeeman Institute, Institute of Physics, University of Amsterdam, Science Park 904, 1098 XH Amsterdam, The Netherlands}
\author{Derek R. Olson}
\author{Anthony P. Austin}
\affiliation{Naval Postgraduate School, Monterey, California 93943, USA}
\author{Joshua A. Dijksman}
\affiliation{Van der Waals-Zeeman Institute, Institute of Physics, University of Amsterdam, Science Park 904, 1098 XH Amsterdam, The Netherlands}
\author{Brian P. Tighe}
\affiliation{Process and Energy Laboratory, Delft University of Technology, Leeghwaterstraat 39, 2628 CB Delft, Netherlands}
\author{Abram H. Clark}
\affiliation{Naval Postgraduate School, Monterey, California 93943, USA}

\begin{abstract}

We use particle-based numerical simulations to address a longstanding question regarding the origins of the linear frequency dependence of attenuation in fluid-saturated granular media.
We study both the acoustic modes and wave propagation in damped, disordered particle packings.
We calculate the damped vibrational modes of packings as a function of frequency, pressure, and grain-contact dissipation. The spatial structure and dissipation of these modes show a clear transition at a pressure-dependent critical frequency from viscous-like continuum behavior to more localized, scattering modes. We also measure how wavespeed and spatial attenuation rate depend on these same parameters. At the same critical frequency, wave propagation also shifts from coherent motion, where attenuation scales quadratically with frequency and linearly with contact damping, to much more incoherent particle-scale motion, where attenuation scales linearly with frequency and sublinearly with contact damping.
All of these features, including the transition frequency, are consistent with a large collection of experimental data, which has not been explained by any framework based on grain-scale physics.
We refer to this approach as ``Jammed-Network Scattering'' (JNS), and propose it as a grain-scale framework for understanding the acoustics of fluid-saturated granular media.

\end{abstract}

\date{\today}

\maketitle	

\section{Introduction \label{sec:Introduction}}

\begin{figure}[t]
    \centering
    \includegraphics[width=.9\columnwidth]{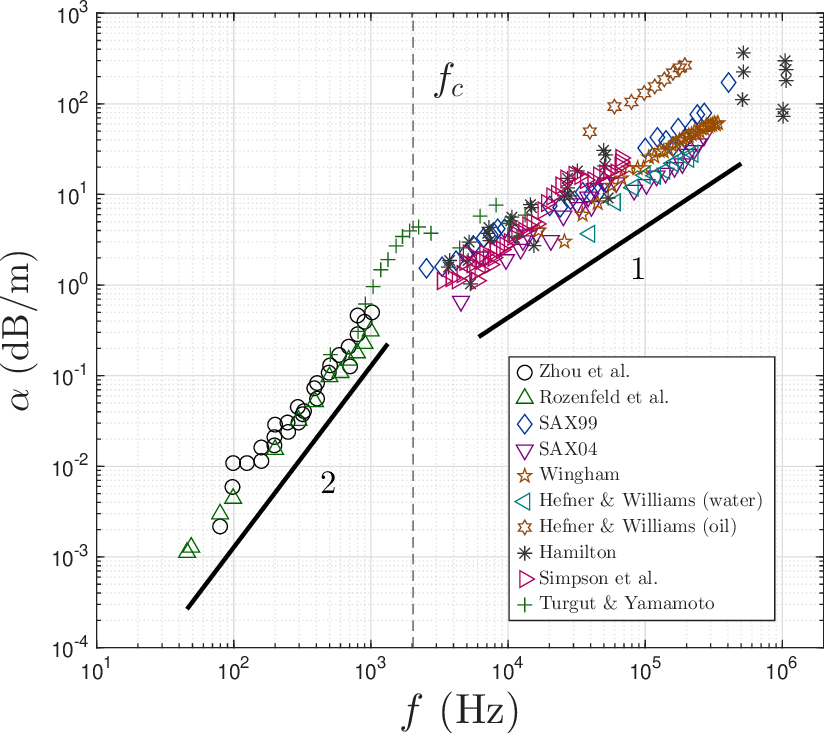}\\[6pt]
    \caption{
     Experimental and field measurements of spatial attenuation coefficient $\alpha$ versus frequency $f$ from multiple marine sediments studies~\cite{Rozenfeld2001,Williams2002a,Hefner2009,zhou_low-frequency_2009, Turgut1990, Wingham1985, hefner2006sound, Hamilton1972, Simpson2003}.
    At low frequencies, the data follow $\alpha \propto f^2$, consistent with viscous dissipation.
    At higher frequencies, the data transition to a nearly linear regime.
    }
    \label{fig:experimentalData}
\end{figure}

Understanding the acoustic properties of fluid-saturated granular materials represents a fundamental challenge in a variety of technology and earth science applications. The acoustic properties of a material are typically characterized by how a phase speed $c$ and spatial attenuation coefficient $\alpha$ depend on frequency $f=\omega/2\pi$. A large collection of data, shown in Fig.~\ref{fig:experimentalData}, from both \textit{in situ} marine sediments (i.e., on the ocean floor) and analogous laboratory measurements of glass beads (with similar diameter and stiffness to sand) immersed in water or silicone oil exhibits anomalous scaling of $\alpha \propto \omega$ for frequencies above $f_c\approx 2$~kHz. These frequencies correspond to wavelengths ranging from a few grain diameters up to roughly a thousand grain diameters~\cite{SAX99_overview,SAX99_environment,zhou_low-frequency_2009, Turgut1990, Wingham1985, hefner2006sound, Hamilton1972,Simpson2003,Rozenfeld2001,Hefner2009,Williams2002a}. At low frequencies, $\alpha \propto \omega^2$ is observed, which agrees with any model based in viscous dissipation, such as the Biot-Stoll model~\cite{biot_generalized_2005,chotiros2017Acoustics}. No model that includes only viscous-like dissipation is known to produce $\alpha \propto \omega$. Additionally, Fig.~\ref{fig:experimentalData} shows that silicone oil, with viscosity of 100 times that of water, produces roughly a tenfold increase in $\alpha$, suggesting a sublinear dependence on viscous effects. The Grain Shear (GS) theory was designed to predict $\alpha \propto \omega$, but required ad hoc grain-grain force laws to obtain this result~\cite{Buckingham1997,buckingham_wave_2000,buckingham_pore-fluid_2007}.

However, no existing acoustic model explicitly incorporates the physics of the disordered granular contact network.
In this paper, we show that (1) linear scaling of $\alpha$ with $\omega$ for $\omega>\omega_c$, (2) sublinear dependence of $\alpha$ on $\gamma$ for $\omega>\omega_c$, and (3) the value of the crossover frequency $f_c$ all arise from the contact network of a granular material near the jamming transition.
Jammed packings are known to support excess low-frequency vibrational (or scattering) modes above a pressure-dependent critical frequency $\hatw > \wc = \hatp^{1/2}$~\cite{Silbert_PRL_2005,Wyart_PRE_2005}, where the circumflex (``hat’’) denotes a dimensionless quantity.

Our results are consistent with the view that scattering of acoustic energy from these structural modes plays an important role in this transition.
We refer to this picture---scattering from the contact network of a jammed packing---as Jammed-Network Scattering (JNS).

To test this framework, we use particle-based numerical simulations to study both the damped vibrational modes and the propagation of compression and shear waves in jammed packings, explicitly including linear dissipation at grain-grain contacts in both normal and tangential directions. We vary the dimensionless pressure $\hatp$, frequency $\hatw$, and grain-contact damping $\hatg$. Rather than using the traditional density of states, we analyze the vibrational spectrum in terms of temporal-attenuation $\hat\beta_i$ versus frequency $\hatw_i$ to establish a direct connection with the spatial attenuation $\hata(\hatw)$ of propagating waves, since $\hata = \hat\beta/\hatc$ for propagation in a given mode. 

We find that, below $\wc$, both temporal and spatial attenuation scale linearly with damping and quadratically with frequency ($\hata \propto \hatg\hatw^2$), consistent with coherent propagation and viscous-like dissipation.
Above $\wc$, propagation becomes highly incoherent, and spatial attenuation transitions to scaling linearly with frequency ($\hata \propto \hatw$) and sublinearly with contact damping ($\hata \propto \hatg^a$ with $a < 1$).
Finally, by substituting typical physical parameters for water-saturated silica sand (or glass beads), we predict the experimental crossover frequency of $f_c \approx 2$~kHz from first principles with no fit parameters.

\begin{figure}
    \centering
    \raggedright (a) \\
    \includegraphics[width=\columnwidth]{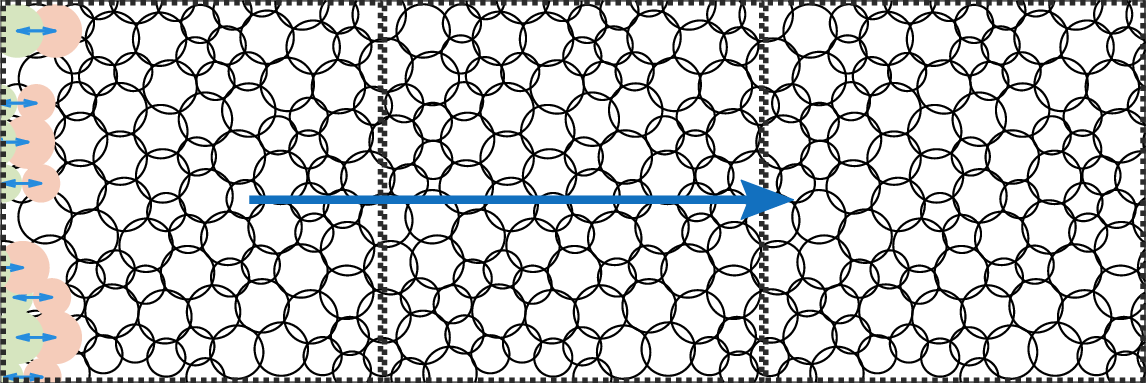}
    \raggedright (b) \\
    \includegraphics[width=\columnwidth]{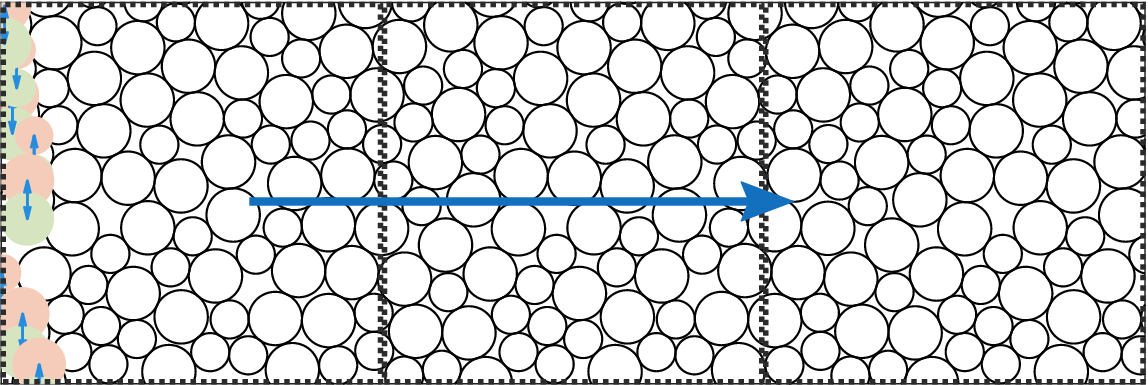}
    \caption{Oscillation simulations for (a) compression waves at large normalized pressure $\hat{P}$ and (b) shear waves at small $\hat{P}$. Particles within one diameter of the wall at $x=0$ are driven sinusoidally with angular frequency $\hat{\omega}$ and amplitude $\hat{P}/100$. 
    }

    \label{fig:packingOscillation}
\end{figure}

In Sec.~\ref{sec:Background} we discuss modeling fluid-saturated granular beds as dissipative mass-spring networks and connect to prior work on the acoustic properties of jammed packings.
In Sec.~\ref{sec:Methods} we describe the computational methods used to obtain our results.
In Sec.~\ref{sec:Results} we show results for the damped acoustic modes, wavespeed, and attenuation for waves propagating through long channels (Fig.~\ref{fig:packingOscillation}).
Section~\ref{sec:Discussion} contains discussion, including a calculation predicting $f_c\approx 2$~kHz from our results. Further details about data analysis are given in Appendices~\ref{app:dft} and \ref{app:data_filtering}.

\section{Background \label{sec:Background}}

Since the bulk modulus of typical sediment grains (e.g., silica) is roughly 20 times that of water, while the mass density is only about twice as large, the internal wave speed inside a grain is roughly 3--4 times the wave speed in water.
For wavelengths near and larger than the grain diameter $d$, the grains therefore respond essentially as rigid bodies, and the wavespeed and attenuation are set by the contact network rather than the grain interiors or the interstitial water.
This supports modeling a fluid-saturated granular bed as a network of rigid bodies connected by dissipative, repulsive contacts: particles of mass $m$ connected by springs of stiffness $\kappa$ and dissipation parameter $\gamma$.
In this framework, it is known that acoustic properties of jammed packings of disks emerge from the contact network structure ~\cite{Ellenbroek2009}.

Existing models for acoustic propagation in fluid-saturated granular media --- including the Biot-Stoll model~\cite{biot_generalized_2005,chotiros2017Acoustics} and Grain Shear theory~\cite{Buckingham1997,buckingham_wave_2000,buckingham_pore-fluid_2007} --- treat the granular backbone as a homogeneous elastic medium.
Neither directly incorporates the disordered contact network structure that arises from the jamming of disordered granular material, which is known to produce anomalous vibrational modes and strongly pressure-dependent mechanical response~\cite{Silbert_PRL_2005,Wyart_PRE_2005,ohern2003jamming,Ellenbroek2009}.

The mechanical and acoustic properties of undamped ($\gamma = 0$) jammed packings have been studied broadly~\cite{ohern2003jamming,van2009jamming,liu2010jamming} and provide a foundation for this study.
Specifically, jammed packings are characterized by the excess packing fraction $\Delta \phi = \phi - \phi_c$ or excess contacts per particle $\Delta z = z - z_c$, both of which scale with the dimensionless pressure $\hatp \sim P/\kappa$ in 2D (or $\hatp \sim Pd/\kappa$ in 3D, not studied here).
Varying $\hatp$ strongly influences energy transport~\cite{vitelli2010attenuation,xu2009energy,Mizuno_modes_2016}, bulk and shear moduli~\cite{ohern2003jamming,Ellenbroek2009}, and vibrational modes~\cite{Silbert_PRL_2005,Wyart_PRE_2005,Mizuno_modes_2016}. A key result is that response of these packings varies dramatically depending on whether the dimensionless frequency $\hatw=\omega\sqrt{m/\kappa}$  is smaller or larger than a critical frequency $\wc = \hatp^{1/2}$. For $\hatw < \wc$, the modes resemble plane waves and the response is elastic continuum-like, while for $\wc < \hatw < 1$ modes are localized and highly non-affine.

For damped packings, relatively little is known about the vibrational modes for varied $\hatp$, or how they relate to dissipation in propagating waves. The dimensionless dissipation parameter is defined as $\hatg\equiv\gamma/\sqrt{\kappa m}$. Recent numerical work investigated dispersion relations in damped granular packings~\cite{Saitoh_sound_char_2023}, focusing on standing waves and their connection to modes, but at fixed, large $\hatp$.
Similarly, Koyama et al.~\cite{koyama2025enhanced} studied direct excitation of modes with damping, but again at fixed large $\hatp$, without propagating waves, and with damping limited to normal dissipation and Stokes damping.
Other work used the proportional damping assumption~\cite{johnson2015density} to simplify the eigenvalue problem, without controlling $\hatp$.
Work in the overdamped limit ($\hatg \gg 1$)~\cite{tighe2010model,tighe2011relaxations} is more relevant to foams than to water-saturated granular media.

A prior work by some of the present authors~\cite{clark2024explicit} showed that propagation through damped jammed packings is viscous-like ($\alpha \propto \gamma\omega^2$) at large $\hatp$, but anomalous ($\alpha \propto \omega^a$, $a \approx 1$) at low $\hatp$.
That work did not disentangle the separate dependences on $\omega$, $\gamma$, and $\hatp$, nor connect the anomalous scaling to the known mode structure of jammed packings.

\section{Methods \label{sec:Methods}}

This section discusses how packings are created and then how they are used in both calculations of the damped acoustic modes as well as in wave propagation simulations. For reference, we provide list of variables used in Table~\ref{tab:all_variables} and of dimensionless variables in Table~\ref{tab:dimensionless_quantities}.

\begin{table}[htpb]
    \centering
    \caption{Summary of core physical, network, and acoustic variables used in the model.}
    \label{tab:all_variables}
    \begin{tabular}{lll}
        \hline\hline
        \textbf{Variable} & \textbf{Name/Description} \\ 
        \hline
        $d$ & Diameter of small particles \\
        $m$ & Mass of a single particle \\
        $\kappa$ & Contact spring stiffness \\
        $\gamma$ & Contact dissipation (damping) parameter \\
        $P$ & Confining pressure \\
        $\bar{U}$ & Mean potential energy per contact \\
        $\phi$ & Packing fraction\\
        $z$ & Contacts per particle\\
        $\rho_{\rm bulk}$ & Bulk density \\
        $B, G$ & Bulk and shear moduli \\
        $\omega$ & Angular frequency ($\omega_d$ for driving frequency) \\
        $c$ & Phase speed ($c_0$ is reference wave speed $d\,\omega_0$) \\
        $\alpha$ & Spatial attenuation coefficient \\
        $k$ & Wavenumber \\
        $\lambda_i$ & Complex eigenvalue of mode $i$ \\
        $\beta_i$ & Temporal decay rate of mode $i$ \\
        $A_x, A_y$ & Longitudinal and transverse oscillation amplitudes \\
        $\phi_x, \phi_y$ & Phase angles referenced to the drive \\
        $N$ & Total number of particles \\
        $L$ & Side length of the simulation box \\
        $D$ & Diameter ratio, $D=1.4$\\
        $F$ & Geometric factor for bulk density \\
        \hline\hline
    \end{tabular}
\end{table}
\begin{table}[htpb]
    \centering
    \caption{Summary of dimensionless quantities.}
    \label{tab:dimensionless_quantities}
    \begin{tabular}{ll}
        \hline\hline
        \textbf{Quantity} & \textbf{Definition} \\ 
        \hline
        $\hat{\omega}$ & $\omega \sqrt{m/\kappa}$ \\
        $\hat{\omega}_c$ & $\hat{P}^{1/2}$ \\
        $\hat{P}$ & $\sqrt{2\bar{U}/(\kappa d^{2})}$ \\
        $\hat{\gamma}$ & $\gamma / \sqrt{\kappa m}$ \\
        $\hat{c}$ & $\frac{c}{d}\sqrt{m/\kappa}$ \\
        $\hat{\alpha}$ & $\alpha d$ \\
        $\hatw_i$ & $\omega_i \sqrt{m/\kappa}$ \\
        $\hat{\beta}_i$ & $\beta_i \sqrt{m/\kappa}$ \\
        \hline\hline
    \end{tabular}
\end{table}

\subsection{Packing generation \label{sec:packing_generation}}

We use discrete element method (DEM) simulations, where Newton's equation of motion is integrated with a velocity-Verlet algorithm under fully periodic boundary conditions, to generate bidisperse frictionless disk packings via isotropic compression. Each packing consists of $N$ disks with equal numbers of small (diameter $d$) and large (diameter $1.4d$) particles of mass $m$, initialized at sparse random positions on a grid with randomized velocities of zero net momentum. The simulation box is square with side length $L$. The resulting square packings can be tiled to create larger systems for wave propagation simulations without additional packing time.

We characterize each packing by its dimensionless pressure $\hat{P}=\sqrt{2\bar{U}/(\kappa d^{2})}$, where $\bar{U}$ is the mean potential energy per particle, and $\kappa$ is the contact stiffness. This form normalizes the confining pressure by the particle elastic modulus, so that $\hat{P}$ measures how compressed the packing is relative to the stiffness of the grains. In our simulations we vary $\hat{P}$ by slowly increasing the packing fraction by increasing $d/L$; however, an equivalent family of states could be achieved experimentally by varying grain stiffness at fixed confining pressure.

Because packing generation only requires damping to dissipate kinetic energy during compression, the contact law used here is simpler than in the wave propagation simulations. During packing we dissipate only the normal component of the relative velocity, i.e. damping is used purely as a numerical relaxation mechanism; the statistics of the final static packings do not depend on this particular damping law or on the damping parameters. The timestep during packing generation is $\Delta t = (\pi/20)\sqrt{M/\kappa}$, and neighbor lists are maintained using a cell lists with width $2Gd$, updated every timestep.

Compression proceeds in two stages. In the first (fast) stage, the box is compressed isotropically at a fractional rate $r_{\text{fast}} = 0.01$ per step until the pressure overshoots $P_{\text{target}}$, at which point the box is expanded back to the previous step. In the second (fine) stage, the box is compressed at a fractional rate $r = P_{\text{target}}$ per step until the kinetic energy falls below $10^{-20}$ and the pressure converges to $P>P_{\text{target}}$. 
After convergence, rattlers (particles with 2 or fewer contacts) are iteratively removed until none remain. The resulting mechanically stable packings are then used both for wave propagation simulations and for computing the damped acoustic modes.

For both the damped modes and the wave propagation simulations we discuss below, we set the mass of each particle to be $m$, despite differences in area.
Thus, the bulk density is not the same as mass density per particle, but can be written as $\rho_{\rm bulk} = mN/L^2$.
Similarly, the packing fraction $\phi$ can be written as

\begin{equation}
   \phi =  \frac{N \frac{\pi}{4}d^2}{L^2}\left[\frac{\left(D^2+1\right) }{2}\right].
\end{equation}
 
where $\frac{\pi}{4}d^2$ is the diameter of small particles and $D$ is the diameter ratio between large and small particles. This allows us to express the bulk density as,

\begin{equation}
   \rho_{\rm bulk} = \frac{m\phi}{d^2} F,
   \label{eqn:rho-bulk}
\end{equation}
where
\begin{equation}
   F =  \frac{4}{\pi}\left[\frac{\left(D^2+1\right) }{2}\right]^{-1} \approx 0.86.
   \label{eqn:F}
\end{equation}

We note that the packing fraction in the packings we use can be well approximated by $\phi = 0.84 + 0.86\,\hat{P} $.

\subsection{Calculating Normal Modes}

To guide an improved analysis for wave propagation, we first consider the normal modes of jammed packings with simple dashpot-like damping (linearly dependent on the velocity difference) between any two particles that are in contact.
Each particle experiences two forces from its neighbors: a restoring spring force proportional to relative displacement, and a viscous drag force proportional to relative velocity.
Combined with its own inertia, this gives the equation of motion for each particle.
The displacements and velocities of all $N$ particles, each with $f$ spatial degrees of freedom, are collected into the $fN$-dimensional vectors $|\mathbf{r}\rangle$ and $|\dot{\mathbf{r}}\rangle$, and the equations of motion for the full packing can be written compactly as
\begin{align}
\mathbf{M}|\ddot{\mathbf{r}}\rangle + \mathbf{\Gamma}|\dot{\mathbf{r}}\rangle + \mathbf{K}|\mathbf{r}\rangle = 0
\end{align}
which is analogous to the scalar damped simple harmonic oscillator, but captures the the dynamics all particles in a linearized equation. 
$\mathbf{M}$, $\mathbf{\Gamma}$, and $\mathbf{K}$ are the mass, damping, and stiffness matrices, encoding inertia, dissipation, and the elastic restoring forces of the contact network, respectively.
Since $m$, $\gamma$, and $\kappa$ are uniform across all particles and contacts for this study, each matrix factors into a scalar times a dimensionless matrix that encodes only the contact trigonometry: $\mathbf{M}=m\mathbf{I}$, $\mathbf{\Gamma}=\gamma \mathbf{\Gamma}'$, and $\mathbf{K}=\kappa \mathbf{K}'$.
Here $\mathbf{\Gamma}'$ encodes which particles are in contact across all degrees of freedom, while $\mathbf{K}'$ captures both the contact network and the overlap of each contact across the degrees of freedom.
Formally, $\mathbf{\Gamma}' = \mathbf{L}\otimes\mathbf{I}_f$ where $\mathbf{L}$ is the $N\times N$ graph Laplacian of the contact network, and $\mathbf{I}_f$ is the $f \times f$ identify matrix, which for each contact damps relative motion in all directions, including transverse the contact normal. 
Finally, $\mathbf{K}'=\mathbf{L}\otimes\mathbf{H}$ where $\mathbf{H}$ is the $f\times f$ contact Hessian.
The key consequence of this factorization is that $m$, $\gamma$, and $\kappa$ do not each independently control the modal structure only their combination $\hatg = \gamma/\sqrt{\kappa m}$ and the contact trigonometry as described next.

Seeking a solution of the form $|\mathbf{r}(t)\rangle=e^{\lambda t}|\mathbf{v}\rangle$ yields the quadratic eigenvalue problem (QEP)
\begin{equation}
\left(\lambda^2 m \mathbf{I} + \lambda \gamma \mathbf{\Gamma}' + \kappa\mathbf{K}'\right)|\mathbf{v}\rangle = 0.
\end{equation}
Non-dimensionalizing by $\sqrt{\kappa/m}$ reduces the system to a single free parameter $\hat{\gamma}$,

\begin{equation}
\left(\hat{\lambda}^2 \, \mathbf{I} +\hat \lambda \hat{\gamma} \,\mathbf{\Gamma}' + \mathbf{K}'\right)|\mathbf{v}\rangle = 0,
\end{equation}

where the dimensionless eigenvalues $\hat{\lambda}_i = -\hat \beta_i \pm i\hatw_i$ have components
\begin{align}
\hat\beta_i &= \frac{\beta_i}{\sqrt{\kappa/m}}, \\
\hatw_i &= \frac{\omega_i}{\sqrt{\kappa/m}},
\end{align}
and the dimensionless damping is
\begin{equation}
\hat{\gamma} = \frac{\gamma}{\sqrt{\kappa m}}.
\end{equation}
This non-dimensionalization shows that the modal structure depends only on $\hat{\gamma}$ and the contact geometry encoded in $\mathbf{\Gamma}'$ and $\mathbf{K}'$, which motivates the parameter sweeps over $\hat{\gamma}$ and $\hat{\omega}$ in the wave propagation simulations described in the following section.

The non-dimensionalized QEP is still nonlinear in $\hat \lambda$, so we introduce the substitution $|\mathbf{w}\rangle = |\mathbf{v}\rangle/\lambda$ and $|\mathbf{z}\rangle = (|\mathbf{v}\rangle, |\mathbf{w}\rangle)^\top$ to reduce it to an equivalent standard eigenvalue problem $\mathbf{A}|\mathbf{z}\rangle = \lambda |\mathbf{z}\rangle$ of twice the dimension for numerical solvers, where

\begin{equation}
\mathbf{A} = \begin{pmatrix} -\gamma \mathbf{\Gamma}' & -\kappa \mathbf{K}' \\ m\mathbf{I} & 0 \end{pmatrix}.
\end{equation}

We solve this problem using standard methods for small $N$ and more advanced contour-integral techniques for low frequencies and large systems, yielding eigenvalues $\hat{\lambda}_i = -\hat{\beta}_i + i\hat{\omega}_i$ and corresponding displacement vectors $|\mathbf{v}_i\rangle$.
For small $N$, we solve this eigenvalue problem using Francis's implicitly-shifted $QR$ algorithm (\texttt{eig} in MATLAB). This approach yields the complete set of modes but rapidly becomes computationally expensive for large systems, as its computational complexity scales as $O(N^3)$. For larger $N$, we use a contour integral projection method to compute only the low-frequency modes of interest \cite{Goe99,Pol09,SS03}, based on the identity
\begin{equation}\label{EQN:DunfordTaylorIntegral}
\mathbf{P} = \frac{1}{2\pi i}\int_\mathcal{C} (z\mathbf{I} - \mathbf{A})^{-1} \, dz,
\end{equation}
where $\mathcal{C}$ is a simple, closed, piecewise-smooth contour in $\mathbb{C}$ enclosing the target eigenvalues and $\mathbf{P}$ is the spectral projector onto the associated eigenspace of $\mathbf{A}$.

In practice, $\mathcal{C}$ is a rectangle enclosing eigenvalues of $\mathbf{A}$ with dimensionless frequency $|\hat{\omega}| < 0.1$. We estimate the bounds of the rectangle using a Krylov--Schur method \cite{Ste2001} (\texttt{eigs} in MATLAB) to approximate the extremal eigenvalues of $\mathbf{A}$ in $\mathbb{C}$. We then compute an approximate orthonormal basis $\mathbf{Q}$ for the target eigenspace by discretizing Eq.~\eqref{EQN:DunfordTaylorIntegral} via Gauss--Legendre quadrature, applying $\mathbf{P}$ to a sufficiently large random set of vectors, and orthonormalizing the results. Finally, the modes of interest are approximated by solving the eigenvalue problem for $\mathbf{Q}^\dagger \mathbf{A} \mathbf{Q}$, which is much smaller than $\mathbf{A}$, using the $QR$ algorithm. Accuracy is improved if needed by repeating this procedure in a subspace iteration.

\subsection{Wave propagation simulations}

For wave propagation DEM simulations, we generate large channels using tiled periodic packings with $N = 100$ (approximately $10\times10$ particles), $400$ ($20\times20$), and $1600$ ($40\times40$) particles per tile. Throughout the simulations, contacts between neighbors are permanent and particle interactions are restricted to linear springs anchored to their initial separations, so no contact formation or breaking occurs. We set the driving amplitude of oscillations (normalized by particle size) to 1\% of the dimensionless pressure $\hatp$.

Inter-particle forces follow a spring-dashpot contact law. The total force on particle $n$ due to particle $m$ is
\begin{equation}
    \mathbf{F}_{nm} = -\kappa\left(\frac{D_{nm}}{d_{nm}} - 1\right)\mathbf{r}_{nm} - \gamma(\mathbf{v}_n - \mathbf{v}_m),
\end{equation}
where $D_{nm} = (D_n + D_m)/2$ is the mean contact diameter, $d_{nm} = |\mathbf{r}_{nm}|$ is the center-to-center distance, and $\gamma$ is the damping coefficient. The damping term acts on the full relative velocity vector rather than only its normal component; as discussed in Sec.~\ref{sec:packing_generation}, this distinction does not affect the static packings from which the simulations begin.

Compressional or shear waves are driven by oscillating all particles that overlap with the $x = 0$ boundary with amplitude $A_{\mathrm{d}}$ and frequency $\omega_{\mathrm{d}}$. For compression, the wall displacement follows $\mathbf{r}_{\mathrm{d}}(t) = A_{\mathrm{d}}\sin(\omega_{\mathrm{d}} t)\,\hat{x}$; for shear, $\mathbf{r}_{\mathrm{d}}(t) = A_{\mathrm{d}}\sin(\omega_{\mathrm{d}} t)\,\hat{y}$. 
In both cases the propagating wavevector is $\mathbf{k} = k\hat{x}$, and particle motion is decomposed into longitudinal ($A_\parallel\hat{x}$) and transverse ($A_\perp\hat{y}$) components. The drive amplitude $A_{\mathrm{d}}$ is two orders of magnitude smaller than the packing pressure, and each simulation runs until the wave has reached approximately 90\% of the channel length. All tile sizes gave consistent results; data shown uses $N = 400$ tiles unless otherwise stated.

We model the steady-state displacement of a particle at the initial equilibrium position $x_0$ from the oscillating boundary as

\begin{equation}
    \mathbf{u}(x_0, t) = \Re\!\left\{\left[A_x(x_0)e^{i\phi_x},\, A_y(x_0)e^{i\phi_y}\right]e^{i\omega_{\mathrm{d}} t}\right\},
\end{equation}
where $A_i(x_0)$ and $\phi_i$ are the amplitude and phase at distance $x_0$ referenced to the drive.
For a coherent plane wave, the amplitudes take the form $A_i(x_0) = A_0\,e^{-\alpha x_0}$ and the phases vary linearly as $\phi_i(x_0) = k x_0 + \phi_0$, so that the displacement field reduces to
\begin{equation}
    \mathbf{u}(x_0,t) \propto \Re\!\left\{ e^{-\alpha x_0}\, e^{i(k x_0 - \omega_{\mathrm{d}} t)} \right\},
\end{equation}
with spatial attenuation coefficient $\alpha$ and wavenumber $k$.
Here $x_0$ is the equilibrium (unperturbed) position of each particle; because the drive amplitude is two orders of magnitude smaller than the packing pressure, mean displacements from equilibrium are confirmed to be negligible in initial measurements and no mean subtraction is applied. The wavespeed is then $c=\omega/k$, and then dimensionless wavespeed is $\hat{c} = c/c_0$, with $c_0=d\sqrt{\kappa/m}$.

\section{Results \label{sec:Results}}
\subsection{Damped Modes}

Previous work on undamped jammed packings has established the presence of excess low-frequency vibrational modes, sometimes called a boson peak. Deviation from continuum elastic response is quantified by comparing the vibrational density of states with Debye scaling~\cite{Silbert_PRL_2005,Wyart_PRE_2005}.
In the present case, where viscous damping is added, each normal mode has a complex eigenvalue $\hat\lambda_i = -\hat \beta_i \pm i\hat\omega_i$.
Here we will focus on the shape of the curve traced out by $(\hat\beta_i,\hat\omega_i)$, rather than on the density of states, to compare with the continuum elastic response.
This also allows direct comparison to the spatial attenuation $\hata(\hat\omega)$ of propagating waves.

\begin{figure}
    \includegraphics[trim=0mm 0mm 0mm 0mm, clip,width=\columnwidth]{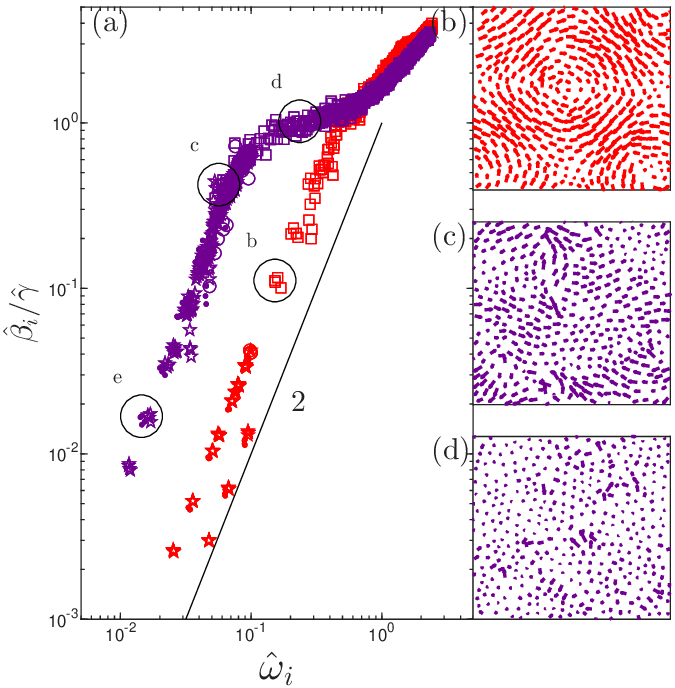}
    \includegraphics[trim=0mm 0mm 0mm 0mm, clip,width=\columnwidth]{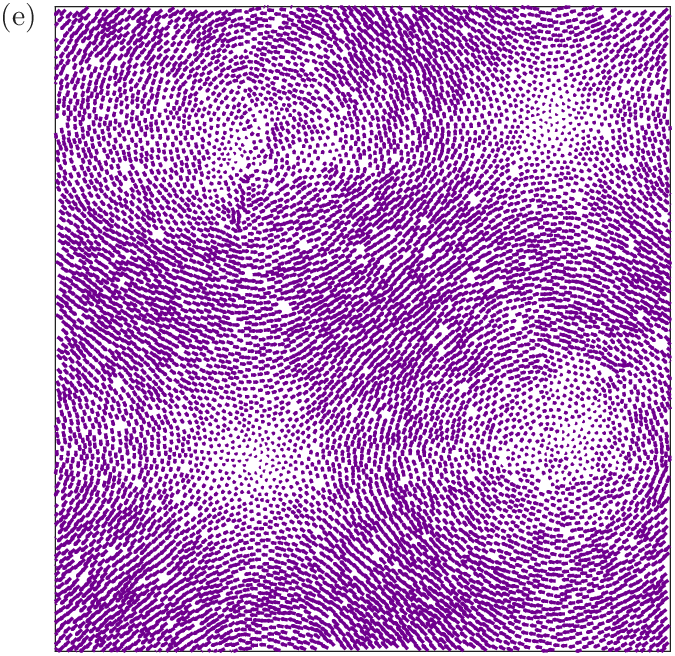}
    \caption{
    (a) A scatterplot of $\hat \beta_i/\hatg$ versus $\hatw_i$ for $\hatg=0.001$, with $\hatp=0.1$ (red symbols) and $\hatp=0.005$ (purple symbols).
    Different symbols shapes represent $N=400$ (squares), 1000 (open circles), 9000 (dots), and 16000 (stars).
    Data for $N=400$ are computed with standard methods and all others use the contour integral projection method in Eq.~\eqref{EQN:DunfordTaylorIntegral}, searching only for eigenvalues with $\hatw_i<0.1$.
    Visualizations of modes are shown in (b-e), corresponding to labels on panel (a), with (b,e) plane-wave-type modes for $N=400$ and 16000 (respectively), (d) a scattering mode, and (c) a mode near the transition.
    }
    \label{fig:eigen_intro}
\end{figure}

Figure~\ref{fig:eigen_intro}(a) shows a scatter plot of the dimensionless decay rate $\hat \beta_i$ versus the dimensionless frequency $\hatw_i$ for $\hatg=0.001$ and two different dimensionless pressures, $\hatp=0.1$ and $\hatp=0.005$.
The different symbol shapes represent different particle number ($N=400$, 1000, 9000, and 16000), emphasizing the agreement between the standard and more advanced methods for computing these eigenvalues. For $\hatp=0.1$, we find $\hat \beta_i\propto {\hatw_i}^2$ for $\hatw_i\lesssim 1$.
These are plane-wave-type modes, and a typical mode from this regime for $N=400$ is shown in Fig.~\ref{fig:eigen_intro}(b).

\begin{figure*}[t]
    \includegraphics[trim=0mm 0mm 0mm 0mm, clip,width=\textwidth]{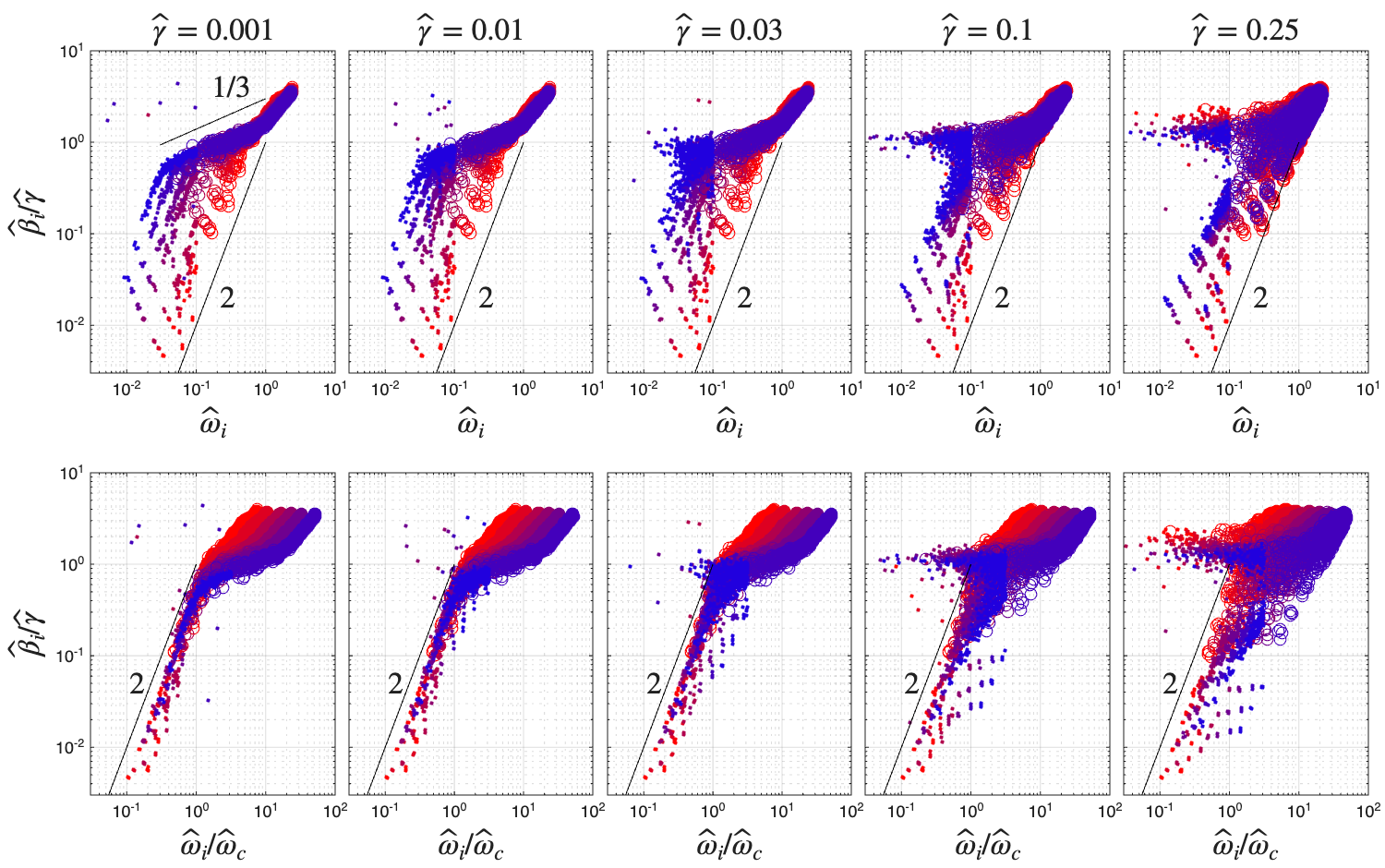}
    \caption{
    Top row: scatter plots of $\hat \beta_i/\hatg$ versus $\hatw_i$ for $0.001<\hatg<0.25$.
    Each plot has many different values of $\hatp$ spanning $0.001<\hat P<0.1$ (symbol color) as well as $N=400$ (circles) and 9000 (dots).
    Blue symbols represent $\hatp=0.001$ and red symbols represent $\hatp=0.1$, with color continuously varied in between.
    Bottom row: the same data is plotted, but with the horizontal axis scaled with $\wc=\hatp^{-1/2}$. 
    }
    \label{fig:eigen_all}
\end{figure*}

For lower pressures (e.g., $\hatp=0.005$), we also observe a regime of plane-wave-type modes with $\hat \beta_i\propto {\hatw_i}^2$.
Figure~\ref{fig:eigen_intro}(e) shows a plane-wave-type mode with $N=16000$.
However, these modes are shifted to lower frequencies and only exist for some $\hatw<\wc$ (we show below that $\wc=\hatp^{1/2})$.
In addition, ``scattering modes'' appear at intermediate frequencies, $\wc<\hatw_i<1$; Fig.~\ref{fig:eigen_intro}(d) shows such a mode for $N=400$.
These modes are similar to the scattering modes studied in previous work that occur at low $\hatp$ ~\cite{Silbert_PRL_2005,Wyart_PRE_2005}.
For scattering modes, $\hat \beta_i$ only depends weakly on $\hatw_i$. For $\hatw_i>1$, both curves ($\hatp=0.1$ and 0.005) are similar.

Figure~\ref{fig:eigen_all} shows scatter plots of $\hat{\beta}_i/\hat{\gamma}$ versus $\hat{\omega}$ and $\hat{\omega}\hat{P}^{-1/2}$ for varying $\hat{\gamma}$ and $\hat{P}$.
Viewing the spectrum in this attenuation–frequency plane makes several features apparent. 
First, the normalized attenuation $\hat{\beta}_i/\hat{\gamma}$ is generally consistent for very low $\hatg$.
The plots shown for $\hatg=0.001$ and 0.01 are nearly indistinguishable; $\hatg=0.03$ shows some distortion but is still very similar.
Thus, in the low-damping regime ($\hatg\lesssim0.1$), the temporal attenuation coefficient $\hat \beta_i$ of each mode is proportional to $\hatg$ but the distribution of $\lambda_i$ in the complex plane is otherwise unchanged.
Thus, the dissipation in any individual mode should be proportional to $\hatg$ for low damping until $\hatg \approx 0.1$.
Additionally, if propagating waves are associated with a fixed frequency band of modes near the driving frequency, then $\hata\propto\hatg$ is expected.
If propagating waves include significant scattering, where many modes are excited, then other behavior may be observed.

Additionally, for low $\hatg$, the defining features of the distribution of the $(\hat \beta_i/\hatg,\hatw_i)$ pairs are set by $\hatw_i = \wc$ and $\hatw_i = 1$.
For the low frequency modes with $\hatw_i<\wc$, $\hat\beta_i\propto {\hatw_i}^2$.
For the scattering modes with $\wc<\hatw_i<1$, $\hat \beta_i/\hatg\sim {\hatw_i}^a$, where $a < 1$.
All data with $\hatw_i>\wc$ lies on the same curve, regardless of the pressure.
The bottom row of Fig.~\ref{fig:eigen_all} shows $\hat \beta_i / \hatg$ plotted versus $\hatw_i/\wc$, demonstrating that the crossover from exponent 2 to exponent $a$ occurs at $\wc$.
When $\hatw_i>1$, we observe $\hat \beta_i\propto {\hatw_i}$.

Finally, we note that when $\hat{\gamma} \gtrsim 0.1$ the distribution of $( \hat \beta_i/\hatg,\hatw_i)$ pairs is distorted such that the low-frequency modes are pushed to the right (higher $\hatw_i$) and the scattering modes are pushed to the left (lower $\hatw_i$).
This is consistent with the data that we show later (in Fig.~\ref{fig:gammaAtten}), where $\hatg>0.1$ leads to changes in the $\hat \beta_i(\hatg)$ curves. Many of the scattering modes become overdamped (i.e., $\hatw_i=0$) as $\hatg$ is increased further. We leave a complete characterization of the transition to overdamped modes for future study.

\begin{figure}
    \raggedright (a) \\
    \centering
    \includegraphics[trim=0mm 0mm 0mm 0mm, clip, width=\columnwidth]{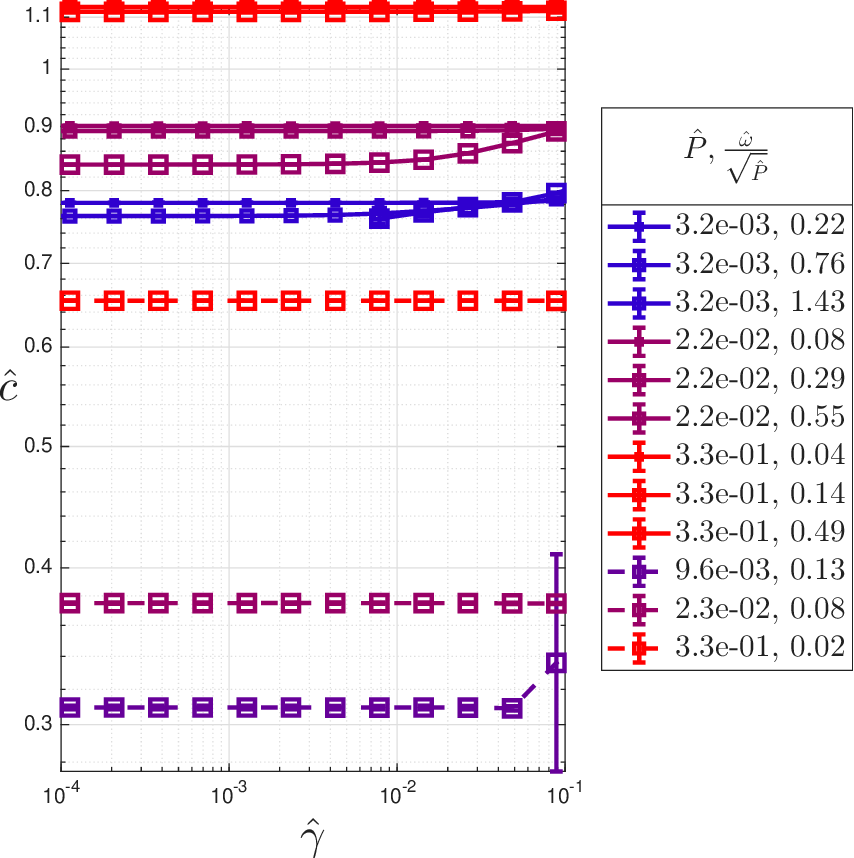}\\
    \raggedright (b)  \hspace{37mm} (c) \\
    \includegraphics[trim=0mm 0mm 0mm 0mm, clip,width=0.47\columnwidth]{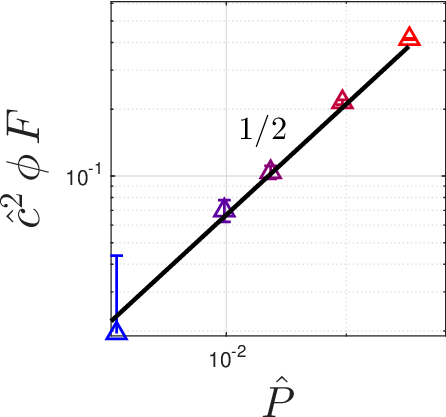}
    \includegraphics[trim=0mm 0mm 0mm 0mm, clip,width=0.47\columnwidth]{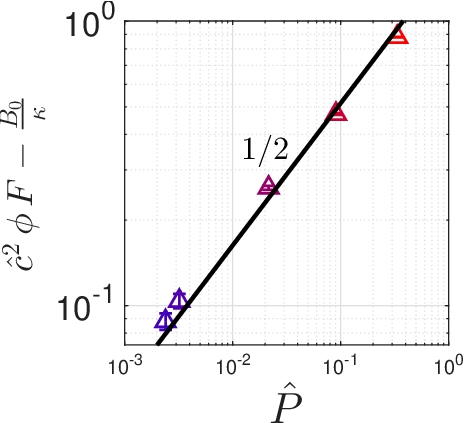}
    \caption{(a) Scaled wavespeed $\hat{c}$ versus $\hat{\gamma}$ for shear (dotted lines) and compression (solid lines) waves in the low-frequency, low-damping limit ($\hat{\omega} < \hat{\omega}_c$, $\hat{\gamma} \to 0$), are mostly insensitive to damping, where higher damping region increases wavespeed (b) Fitted shear wavespeed as a function of $\hat{P}$ with power-law fit and $F$ parameter as detailed in Eq.~\eqref{eqn:wavespeedModulus}; the slope recovers the established $P^{1/2}$ scaling of jammed packings~\cite{Ellenbroek2009}.
    (c) Fitted compression wavespeed versus $\hat{P}$, with $B_0/\kappa = 0.34$ fixed from~\cite{Ellenbroek2009} and the $P^{1/2}$ slope fitted freely.
    Fitted exponents and prefactors are in close agreement with prior work, confirming that our simulations reproduce the known elastic response before damping and scattering effects are introduced.
    For shear waves, the lowest-pressure curve ($\hat{P} = 3.3\times10^{-3}$) is omitted because even our smallest driving frequency excited strongly scattered waves, preventing a reliable wavespeed estimate.
    This collapses curves of different $\hatp$ at fixed $\hatg$ below $\wc$, yielding $\hata \propto \hatg^b\hatw^2\hatp^{1/4}$.
    }

    \label{fig:wavespeedCombined}
\end{figure}

\subsection{Wave Speed}

Before examining spatial attenuation from numerical simulations, we first benchmark our simulations against known elastic behavior by measuring wave speed.
Experimental data from~\cite{zhou_low-frequency_2009} shows that wave speed does not depend strongly on frequency; however, increases a few percent at around the same frequency where the attenuation coefficient changes from quadratic to linear. In this section we find little dependence of $\hatc$ on $\hatw$, but we fail to reproduce the slight increase observed in experimental data. We also show that the dependence of $\hatc$ on $\hatg$ is weak. We show that wave speed scales with $\hatp$ in a way that is consistent with previous work on the elastic modulus of packings near jamming.

Continuum mechanics predicts that wavespeed depends on bulk modulus $B$, shear modulus $G$, and bulk density $\rho_\text{bulk}$, defined in Sec.~\ref{sec:Methods}.
In two dimensions,

\begin{equation}
    c = \left[\frac{B + G}{\rho_\text{bulk}}\right]^{1/2},
    \qquad
    \begin{cases}
        B = B_0 + B_1\,P^{1/2} \\
        G = G_0\,P^{1/2}.
    \end{cases}
\end{equation}

Using Eqs.~\eqref{eqn:F} and ~\eqref{eqn:rho-bulk}, we can use these physical parameters to express dimensionless wavespeed as a function of pressure as

\begin{equation}
    \hat{c}^2\,\phi \, F= \frac{B_0}{\kappa} + \frac{\left(B_1 + G_0\right)}{\kappa}P^{1/2}.
     \label{eqn:wavespeedModulus}
\end{equation}

 Figure~\ref{fig:wavespeedCombined}(a) shows curves for $\hatc(\hatg)$ for different combinations of $\hatp$ and $\hatw$.
 These data show that $\hatc$ is largely insensitive to $\hat{\gamma}$ in the low-damping regime.
 We note data for $\hatp=0.0032$ that span $\hatw/\wc<1$ and $\hatw/\wc>1$ with no meaningful difference.
 Figure~\ref{fig:wavespeedCombined}(b) and (c) show the values $\hatc$ for shear waves, in panel (b), and compression waves, in panel (c), of the low-$\hatg$ limit plotted as a function of $\hatp$.
For shear waves, in the low-damping limit ($\hat{\gamma} \to 0$), the shear wavespeed recovers $\hat{c}^2 \propto P^{1/2}$, consistent with the well-studied pressure scaling of the shear modulus in jammed packings~\cite{Ellenbroek2009}. The bulk modulus contribution vanishes for shear waves, as expected, and fitting yields $G_0/\kappa = 0.723$, in close agreement with prior work~\cite{Ellenbroek2009}.

For compression waves, our data are consistent with a zero-pressure bulk modulus $B_0/\kappa = 0.34$ from Ellenbroek et al.~\cite{Ellenbroek2009}.
The fitted slope recovers the expected $P^{1/2}$-scaling with $(G_0 + B_1)/\kappa = 1.55$, as shown in Fig.~\ref{fig:wavespeedCombined}(c), also in agreement with~\cite{Ellenbroek2009}.
This implies that $B_1=.827$ when taken with $G_0$ from shear results.
Like shear, compression wavespeed is largely insensitive to damping within the low-damping regime.

The recovery of established pressure-scaling exponents and modulus values from both wave types confirms that our simulations reproduce the elastic response of jammed packings. However, we note that slight increase of $c$ at $\omega=\omega_c$, which is observed in experimental data~\cite{zhou_low-frequency_2009}, does not obviously appear in our data.

\subsection{Spatial Attenuation - Damping Dependence}

In this section, we show data for $\hata$ for varied $\hatg$, $\hatp$, and $\hatw$, focusing on how $\hata$ varies with $\hatg$.
Our primary finding is that attenuation of waves propagating through damped granular packings is consistent with viscous-like dissipation at $\hatw<\wc$. For $\hatw>\wc$, attenuation does not obey a viscous-like scaling and appears to be dominated by scattering from anomalous modes.

\begin{figure}
    \raggedright (a) \\
    \centering
    \includegraphics[trim=0mm 0mm 0mm 0mm, clip, width=\columnwidth]{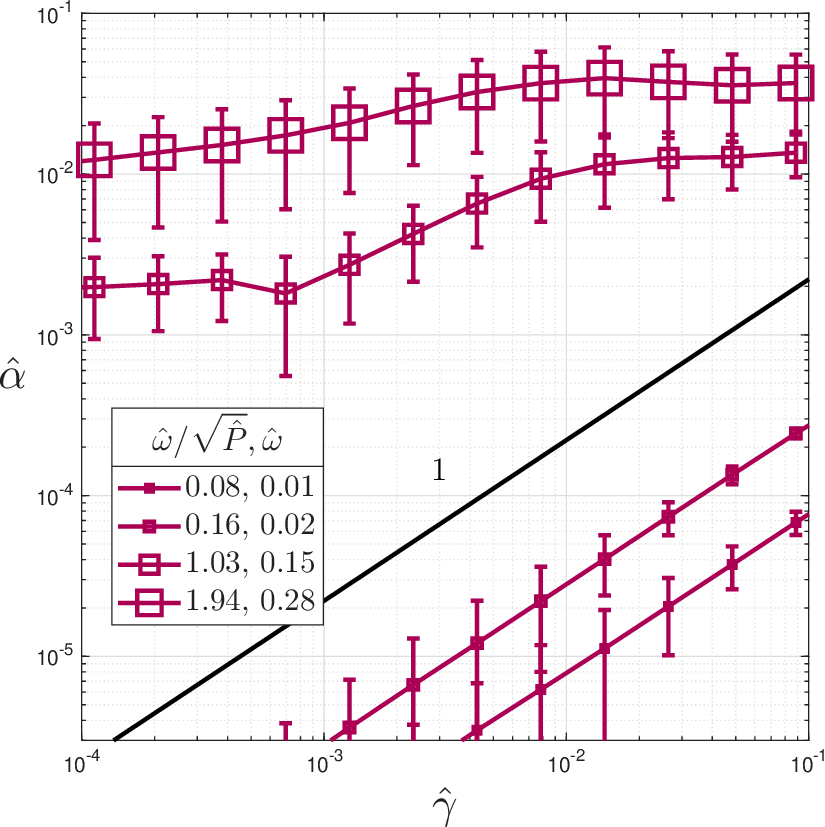}        \raggedright (b)  \\
\includegraphics[trim=0mm 0mm 0mm 0mm, clip, width=\columnwidth]{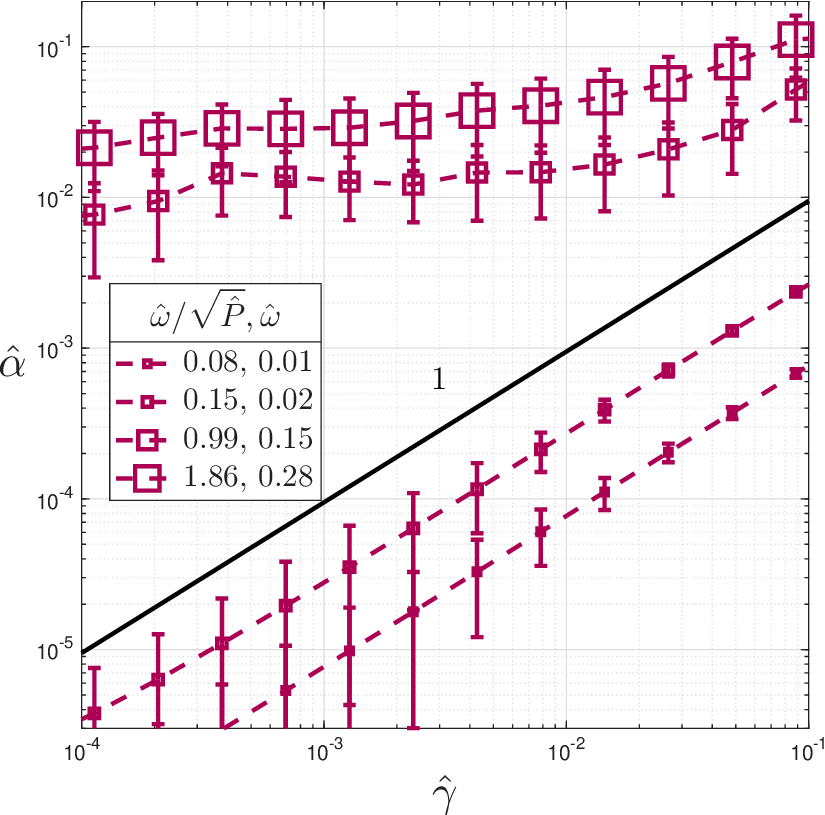}\\
    \caption{
    Attenuation $\hata$ versus $\hatg$ at $\hatp \approx 0.02$ for selected values of $\hatw$ (larger symbols correspond to larger $\hatw$), for (a) compression and (b) shear waves.
    Below $\wc$, curves are evenly spaced and $\hata \propto \hatg$, reflecting viscous-like dissipation.
    Above $\wc$, the $\hatg$-dependence weakens significantly, indicating the onset of scattering-dominated attenuation.
    }
    \label{fig:gammaAtten}
\end{figure}

Figure~\ref{fig:gammaAtten} shows representative curves of $\hata \equiv \alpha d$ versus $\hatg$ for both $\hatw<\wc$ and $\hatw>\wc$ (denoted by symbol size) at moderate pressure $\hatp\approx 0.02$, for compression in panel (a) and shear waves in panel (b).
When $\hatw < \wc$, $\hata \propto \hatg$, indicating viscous-like dissipation, consistent with coupling to plane-wave-like modes where $\hat \beta_i \propto \hatg\hatw_i^2$ as shown in Fig.~\ref{fig:eigen_intro}(b,c,e).
Additionally, the spacing of the lines is even and (as we show below) obeys $\hata\propto\hatw^2$.
When $\hatw > \wc$, $\hata$ is substantially larger and its dependence on $\hatg$ becomes much weaker, indicating that the scattering mechanism introduces a fundamentally different attenuation behavior than viscous-like damping.

\begin{figure}
    \raggedright (a)  \hspace{37mm} (b) \\
    \includegraphics[trim=0mm 0mm 1mm 0mm, clip,width=0.49\columnwidth]{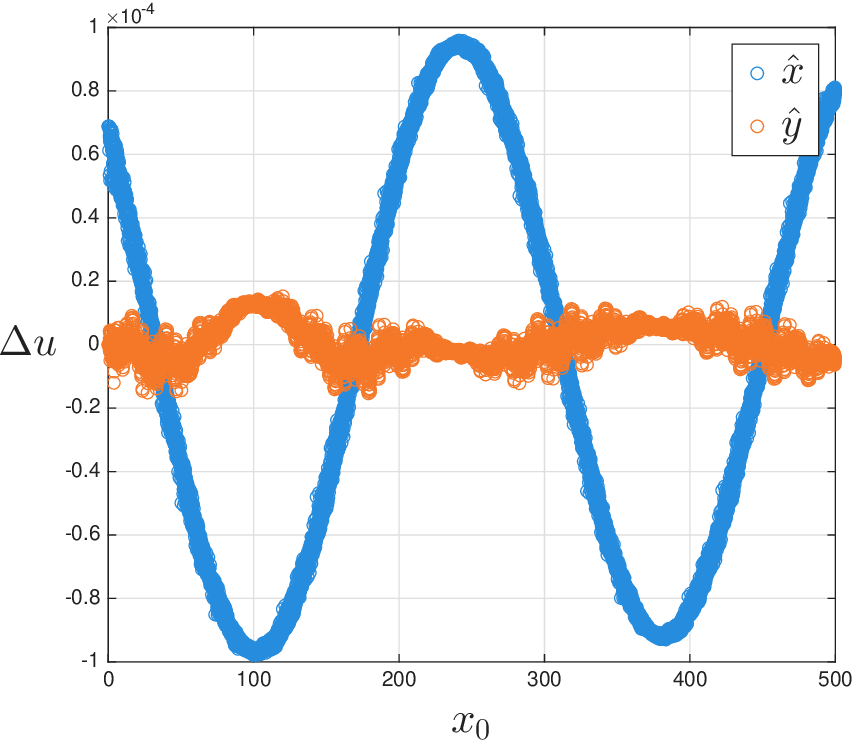}
    \includegraphics[trim=0mm 0mm 1mm 0mm, clip,width=0.49\columnwidth]{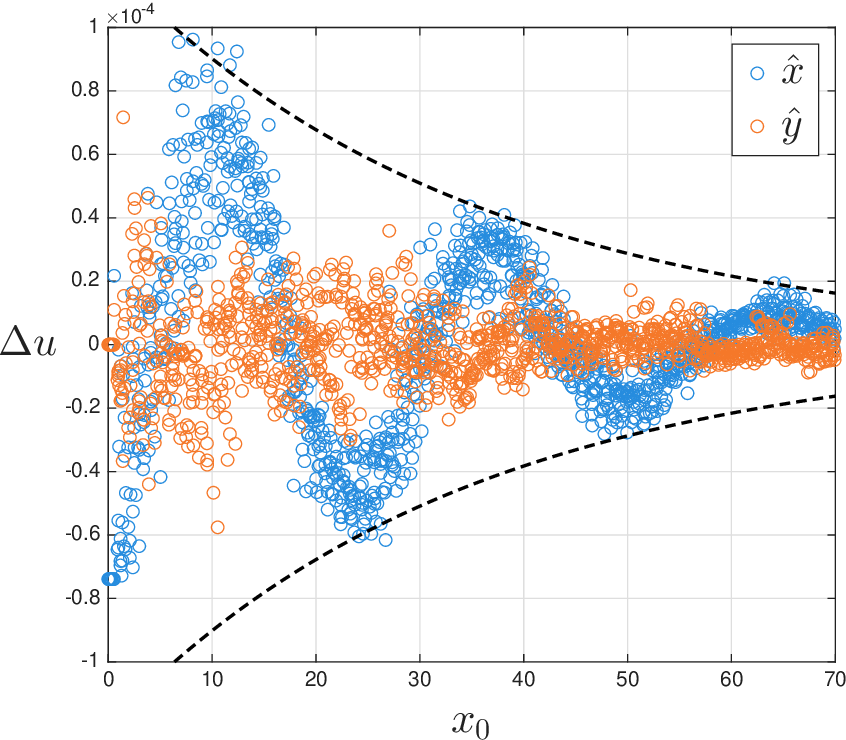}\\
    \raggedright (c)  \hspace{37mm} (d) \\
    \includegraphics[trim=0mm 0mm 1mm 0mm, clip,width=0.49\columnwidth]{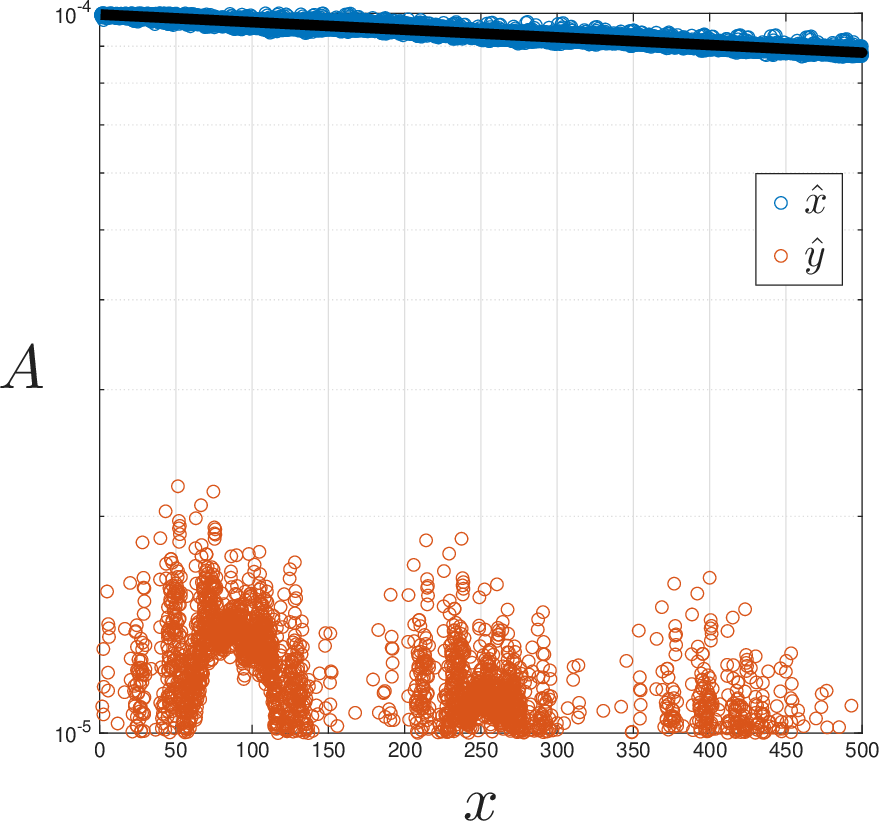}
    \includegraphics[trim=0mm 0mm 1mm 0mm, clip,width=0.49\columnwidth]{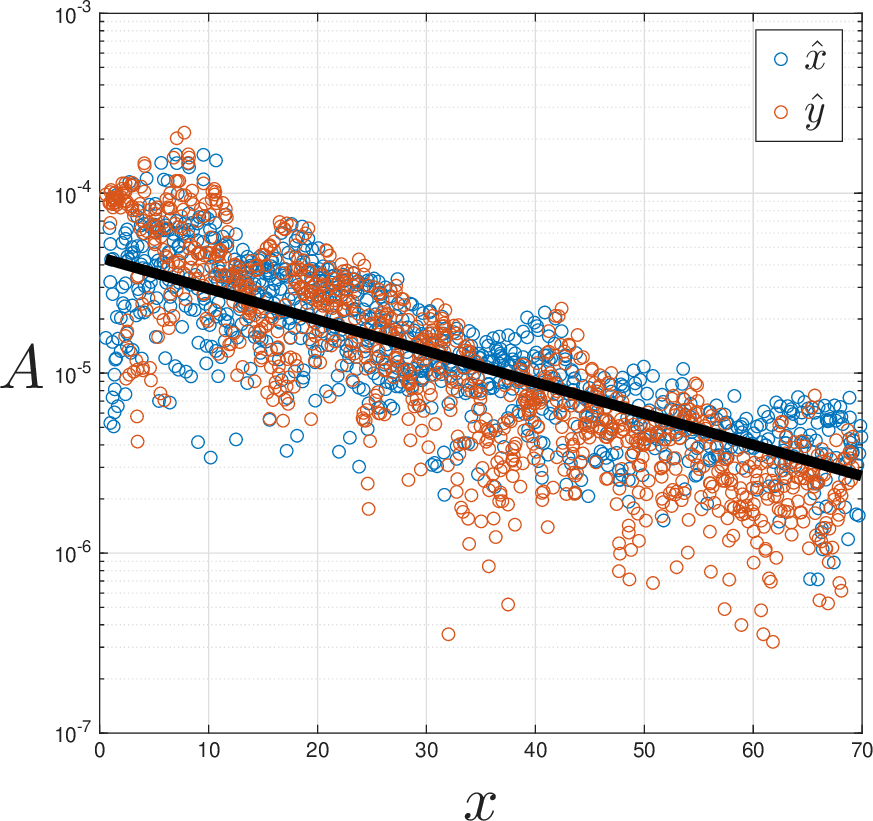}
    \raggedright (e)  \hspace{37mm} (f) \\
    \includegraphics[trim=0mm 0mm 1mm 0mm, clip,width=0.49\columnwidth]{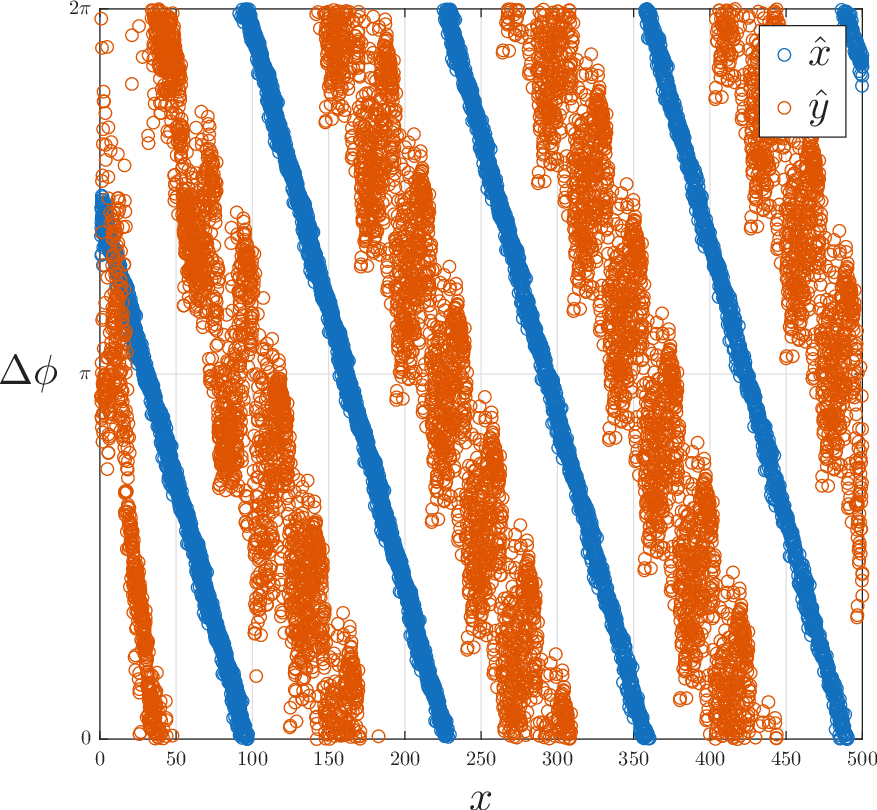}
    \includegraphics[trim=0mm 0mm 1mm 0mm, clip,width=0.49\columnwidth]{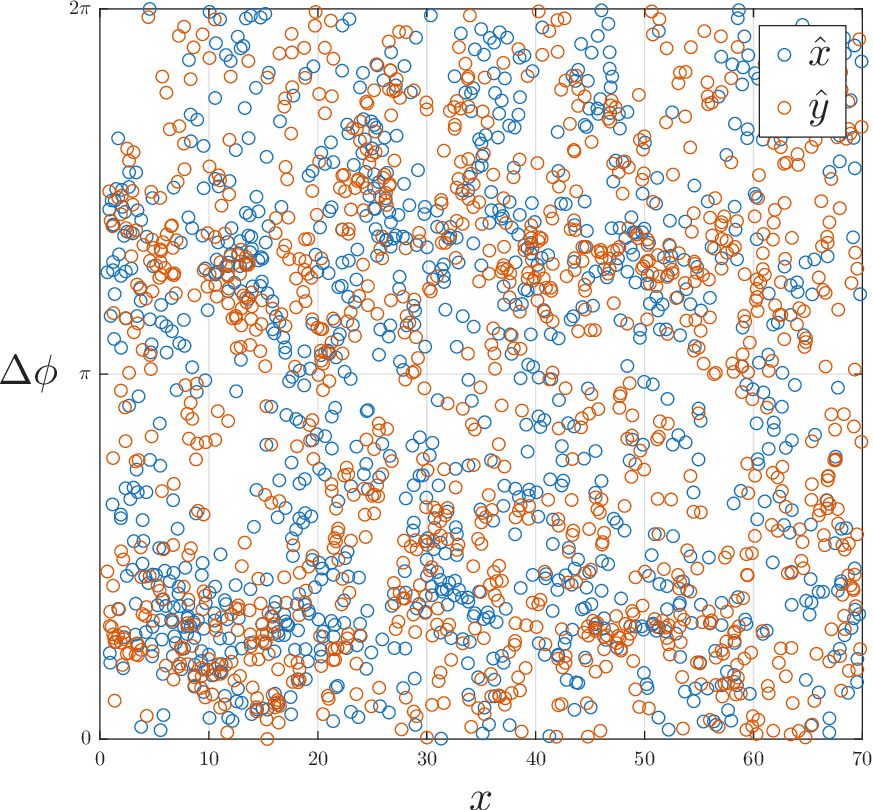}
    \caption{
    Compression wave amplitude and phase data contrasting the viscous regime ($\hat{\omega} < \hat{\omega}_c$, left column, panels a, c, e) and the scattering regime ($\hat{\omega} > \hat{\omega}_c$, right column, panels b, d, and f), both at $\hat{P} \approx 0.02$ and $\hat{\gamma} = 0.1$.
    (a,b) Snapshots of instantaneous particle displacement $A$ versus initial position $x_0$, showing longitudinal ($\hat{x}$, blue) and transverse ($\hat{y}$, orange) components. In the viscous regime (a), longitudinal displacements are coherent and sinusoidal with $A_y \ll A_x$.
    In the scattering regime (b), both components show large scatter and similar magnitudes.
    (c,d) Amplitude data extracted via the on-the-fly DFT described in Appendix~\ref{app:dft}.
    In the viscous regime (c), $A_x$ decays smoothly and $A_y \ll A_x$ throughout the channel.
    In the scattering regime (d), both amplitudes show large scatter and $A_y \sim A_x$, consistent with the disordered scattering modes above $\hat{\omega}_c$.
    (e,f) Wrapped phase $\Delta\phi$ versus $x$ for longitudinal (blue) and transverse (orange) components.
    In the viscous regime (e), the longitudinal phase shifts coherently, while the transverse phase shows early signs of decoherence.
    In the scattering regime (f), both components are highly disordered, reflecting the loss of a well-defined wavevector.
    }
    \label{fig:compAmpPhase}
\end{figure}

\begin{figure}
    \raggedright (a)  \hspace{37mm} (b) \\
    \includegraphics[trim=0mm 0mm 0mm 0mm, clip,width=0.48\columnwidth]{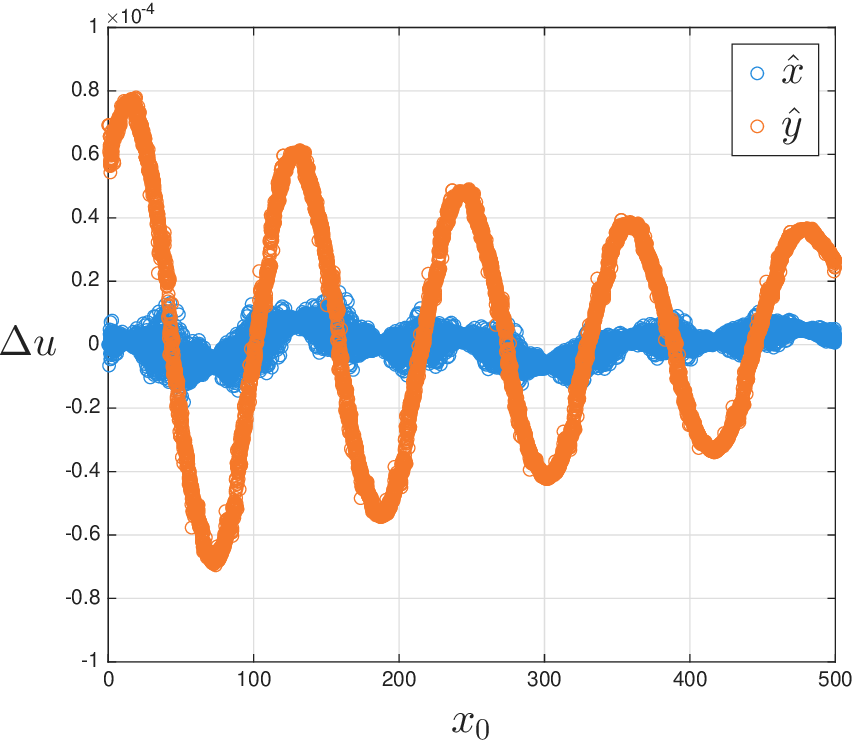}
    \includegraphics[trim=0mm 0mm 145mm 0mm, clip,width=0.48\columnwidth]{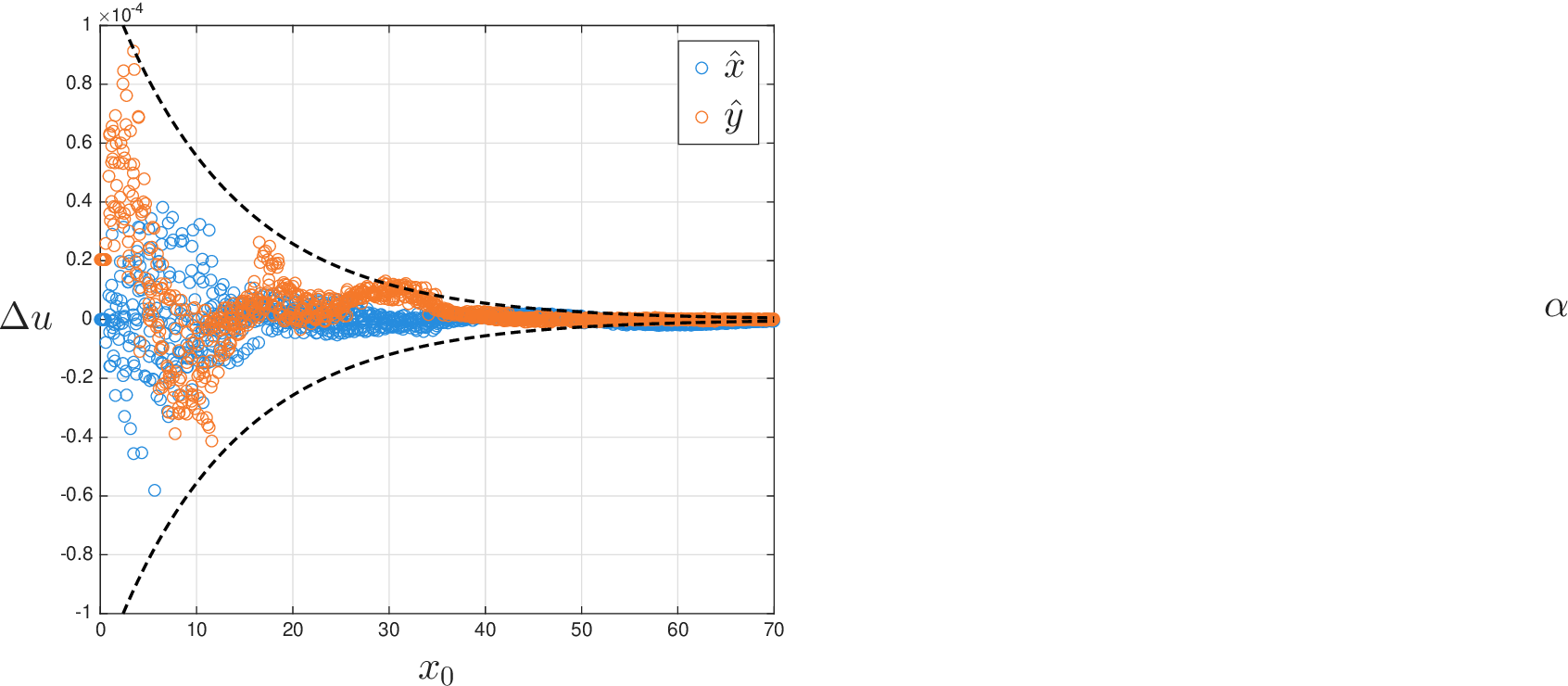}\\
    \raggedright (c)  \hspace{37mm} (d) \\
    \includegraphics[trim=0mm 0mm 1mm 0mm, clip,width=0.49\columnwidth]{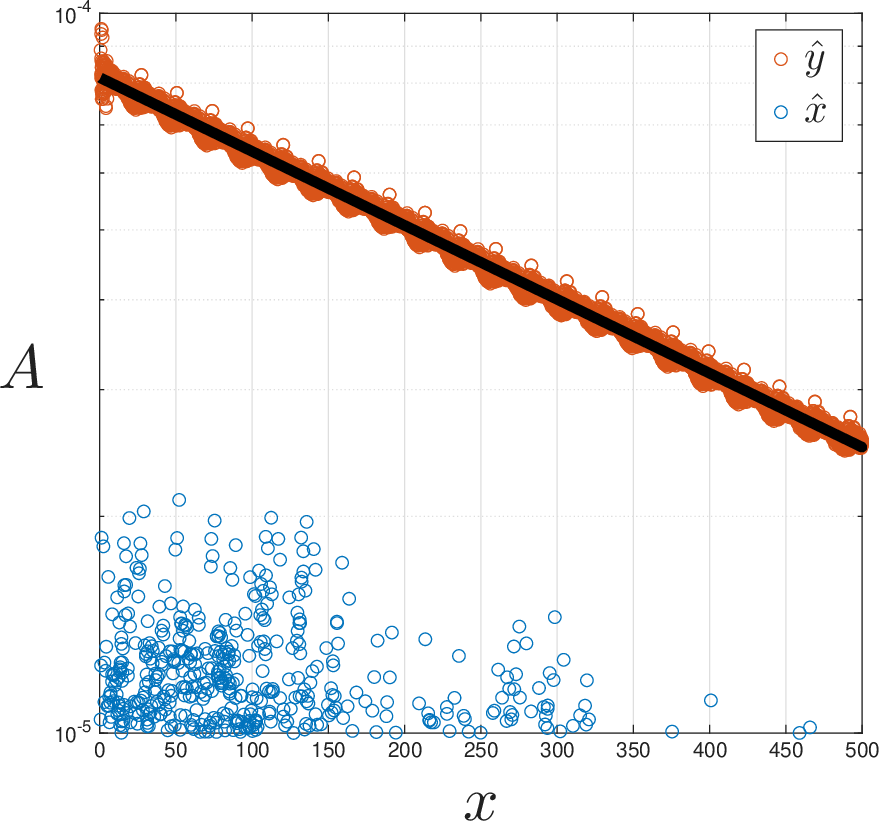}
    \includegraphics[trim=0mm 0mm 1mm 0mm, clip,width=0.49\columnwidth]{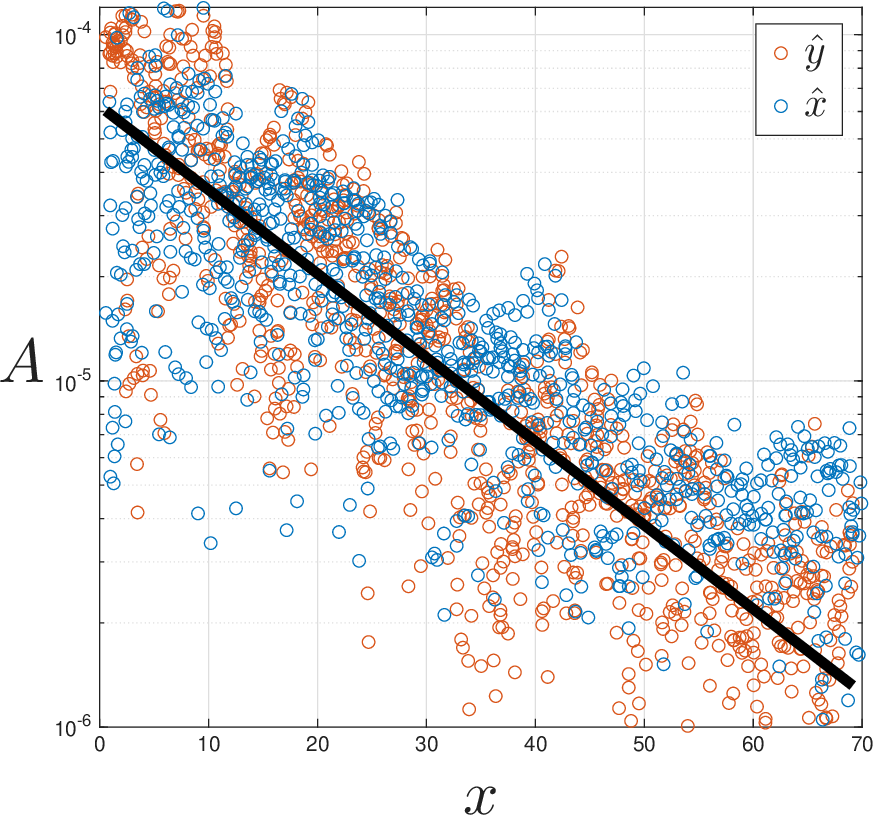}
    \raggedright (e)  \hspace{37mm} (f) \\
    \includegraphics[trim=0mm 0mm 1mm 0mm, clip,width=0.49\columnwidth]{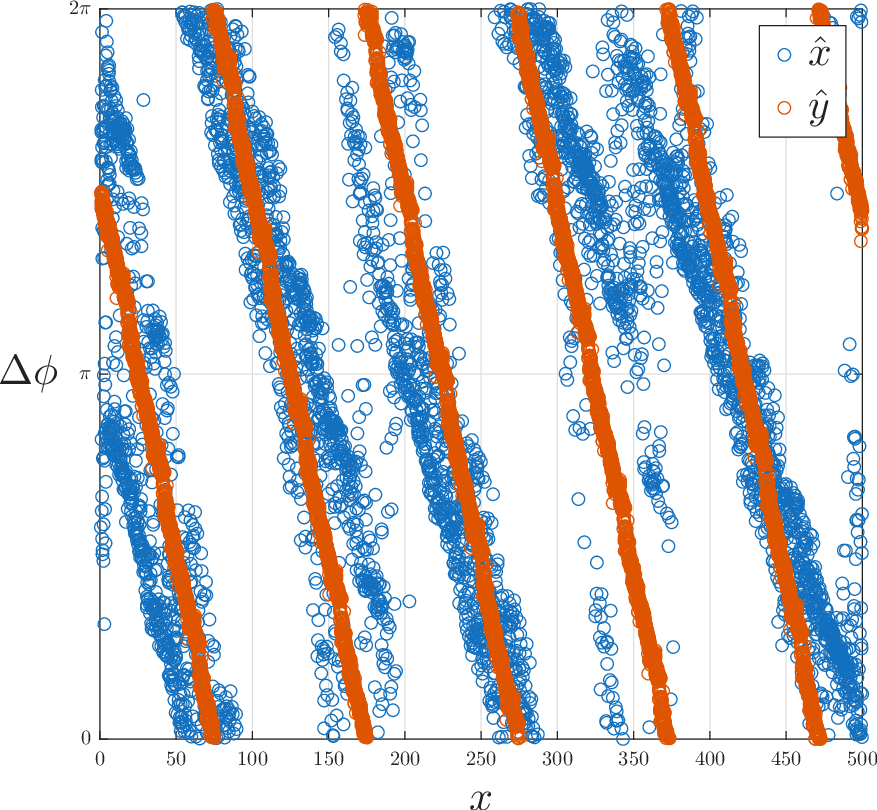}    
    \includegraphics[trim=0mm 0mm 1mm 0mm, clip,width=0.49\columnwidth]{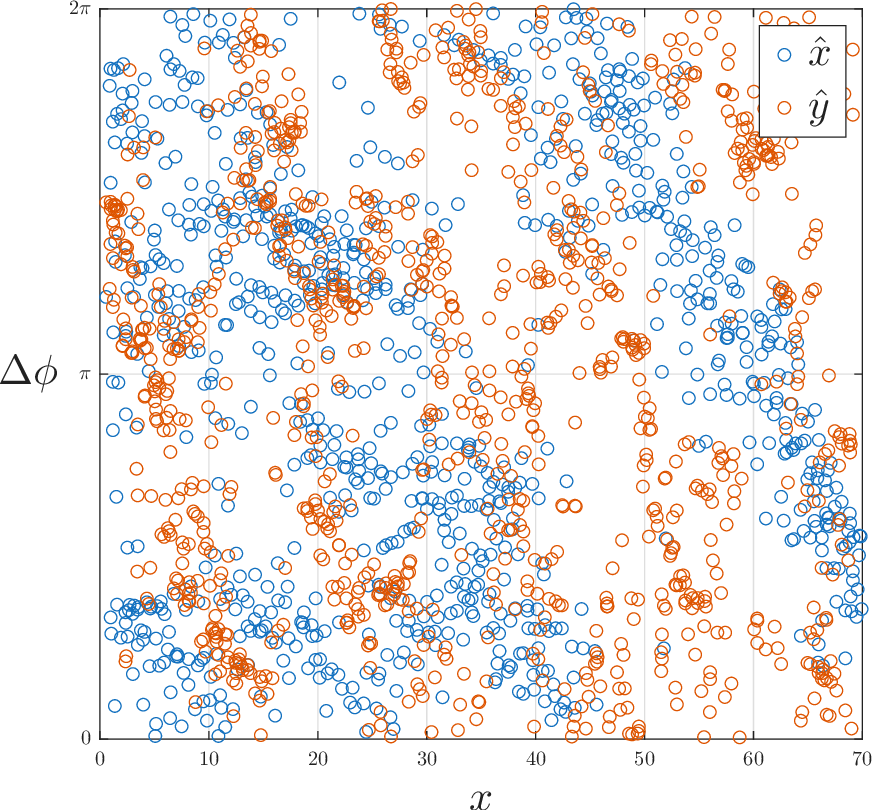}

    \caption{
    Shear wave amplitude and phase data at the same parameters as Fig.~\ref{fig:compAmpPhase} ($\hat{P} \approx 0.02$, $\hat{\gamma} = 0.1$), contrasting the viscous regime ($\hat{\omega} < \hat{\omega}_c$, left column) and scattering regime ($\hat{\omega} > \hat{\omega}_c$, right column).
    In the viscous regime (a,c,e), $A_x \ll A_y$ and the transverse phase shifts coherently, with notably more attenuation than the compression case.
    In the scattering regime (b,d,f), $A_x \sim A_y$ and both phase components are disordered, consistent with the scattering modes above $\hat{\omega}_c$.
    }
    \label{fig:ampPhaseShear}
\end{figure}

Figures~\ref{fig:compAmpPhase} and~\ref{fig:ampPhaseShear} show particle-level data for propagating waves, including a snapshot of the displacements $\Delta u$ during wave propagation, the amplitudes of longitudinal and transverse particle oscillations, and the phase of each oscillator.
Beginning with Fig.~\ref{fig:compAmpPhase}, which shows data for compression waves, panels (a) and (b) show a snapshot of instantaneous particle displacement as a function of initial position $x_0$ for $\hatw<\wc$ (a) and $\hatw>\wc$ (b).
Blue data represent longitudinal oscillations, which are dominant for compression waves, and orange represents transverse oscillations.
This is also shown in panels (c) and (d), where the steady-state oscillation amplitude $A_i(x_0)$, extracted via the on-the-fly DFT, is shown as a function of position along the channel.
When $\hatw<\wc$, $A_y \ll A_x$, as expected for a coherent compression wave.
When $\hatw>\wc$, the transverse oscillation amplitude becomes comparable to the longitudinal amplitude, $A_y \sim A_x$.
Finally, panels (e) and (f) show the phase extracted from each DFT, showing coherence for $\hatw<\wc$ and a complete loss of coherence for $\hatw>\wc$.
Fig.~\ref{fig:ampPhaseShear} shows the analogous data for shear waves at the same parameters.
These plots demonstrate that propagation with $\hatw > \wc$ is characterized by oscillations perpendicular to the wave polarization with comparable magnitude to those along the polarization direction, as well as large phase differences between neighboring particles.
This is consistent with the disordered, non-plane-wave character of the scattering modes above $\wc$ identified in Fig.~\ref{fig:eigen_intro}(d).

Taken together, the $\hatg$-dependence shown in Fig.~\ref{fig:gammaAtten} and the particle-level amplitude and phase data in Figs.~\ref{fig:compAmpPhase} and~\ref{fig:ampPhaseShear} confirm that attenuation crosses over from a viscous-dominated to a scattering-dominated regime at $\wc$.
This is further quantified in Figs.~\ref{fig:shearAttenuation} and~\ref{fig:compAttenuation}, which show $\hata(\hatw)$ at fixed $\hatg$.
In a real system, $\hatg$ is set by a combination of grain size, fluid viscosity, and intrinsic grain inelasticity (characterized by the restitution coefficient), while $\hatw$ varies with the frequency of the propagating wave.
The linear dashpot model used here subsumes both viscous and inelastic dissipation mechanisms into the single parameter $\hatg$, so the results are general to both physical origins.
In the viscous regime ($\hatw<\wc$), $\hata\propto\hatg\hatw^2$ for both wave types.
In the scattering regime ($\hatw>\wc$), the $\hatg$-dependence weakens: $\hata$ becomes weakly dependent on $\hatg$ for shear waves and nearly independent of $\hatg$ for compression waves.
The scaling of $\hata$ with $\hatp$ and $\hatw$ also changes across $\wc$.

\subsection{Spatial Attenuation - Frequency Dependence}

\begin{figure}[t]
    \centering
    \includegraphics[width=.9\columnwidth]{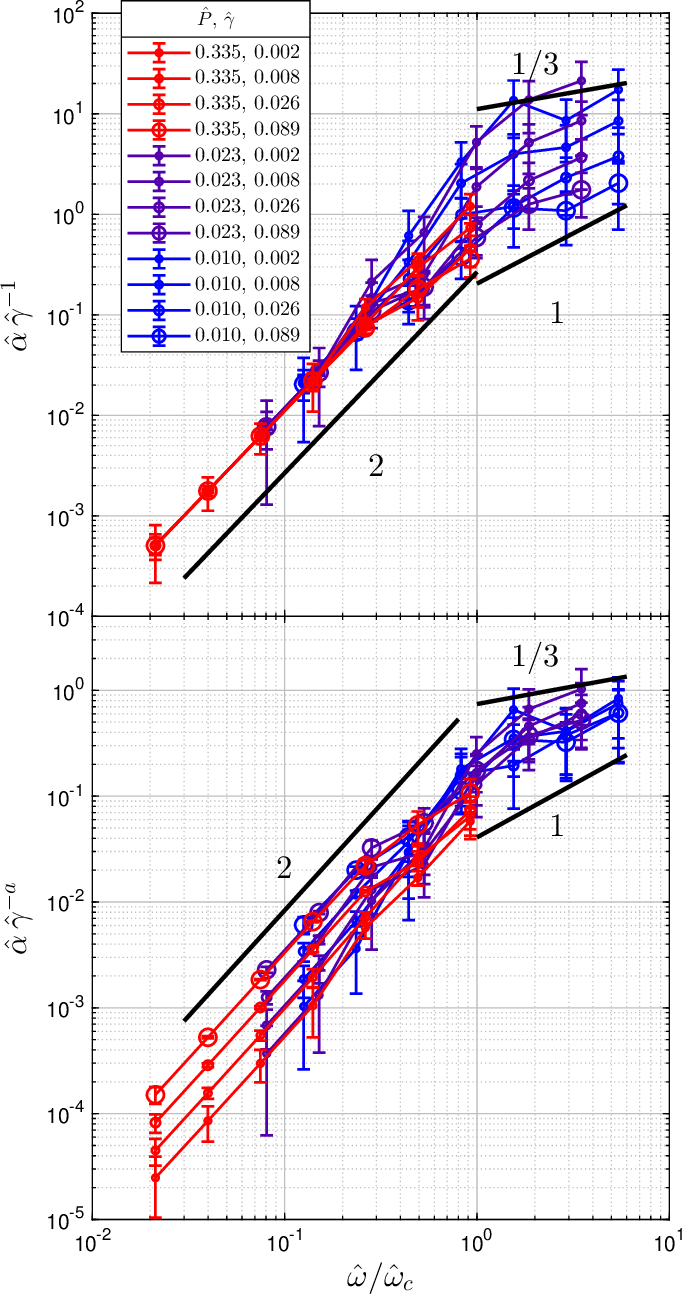}\\[6pt]
    \caption{
    Shear wave attenuation normalized by $\hatg$ (top) and a sublinear coefficient $\hatg^{a}$ (bottom) versus $\hatw/\wc$, for selected values of $\hatp$ and $\hatg$ (see legend; color indicates $\hatp$, red to blue from high to low).
    Below $\wc$, collapsing $\hata/\hatg$ onto a single curve with slope 2 confirms $\hata \propto \hatg\hatw^2$, consistent with the modal scaling in Fig.~\ref{fig:eigen_all} and the $\hatg$-dependence shown in Fig.~\ref{fig:gammaAtten}.
    Above $\wc$, the $\hata/\hatg$ curves spread, indicating that the linear $\hatg$-dependence breaks down; the bracketed reference lines show slopes 1/3 and 1, bounding the observed scaling.
    We rescale with an effective weak power law $\hatg^{a}$ (bottom) to account for the sublinear but non-constant gamma dependence in Fig. \ref{fig:gammaAtten}.
    This scaling improves the collapse above $\wc$, suggesting $\hata \propto \hatg^{a}\hatw^{4/3}$ where $a<1$ in the scattering regime, though the large error bars reflect the reduced fit fidelity inherent to the strongly scattered wave field at these frequencies.
    }
    \label{fig:shearAttenuation}
\end{figure}

Figure~\ref{fig:shearAttenuation} shows $\hata(\hatw)$ for shear waves, normalized by $\hatg$ (top) and $\hatg^{a}$ where $a<1$ (bottom), for a range of $\hatg$ and $\hatp$.
Below $\wc$, the data collapse onto a single curve with slope 2 when normalized by $\hatg$, confirming $\hata \propto \hatg\hatw^2$, consistent with the viscous result of our previous work~\cite{clark2024explicit}.
Above $\wc$, the $\hata/\hatg$ curves spread, indicating the breakdown of linear $\hatg$-dependence; the observed scaling is sublinear in $\hatg$, bounded between slopes of 1/3 and 1. The exponent of $1/3$ is shown for comparison with the modes plots in Fig.~\ref{fig:eigen_all}.
Normalizing by an effective power law $\hatg^{a}$ improves the collapse above $\wc$, suggesting $\hata \propto \hatg^{a}$ with $a < 1$ in the scattering regime, though large error bars from the strongly scattered wave field make a precise exponent difficult to determine.
We note that the $\hatg$-dependence in the scattering regime is more nuanced than a simple power law: as seen in Fig.~\ref{fig:gammaAtten}(b), the curves above $\wc$ not only shift with $\hatg$ but also change shape, indicating that the effective exponent varies across the scattering regime.
The rescaling by $\hatg^{a}$ should therefore be understood as capturing the overall sublinear trend rather than an exact scaling collapse.

\begin{figure}[t]
    \centering
    \includegraphics[width=.9\columnwidth]{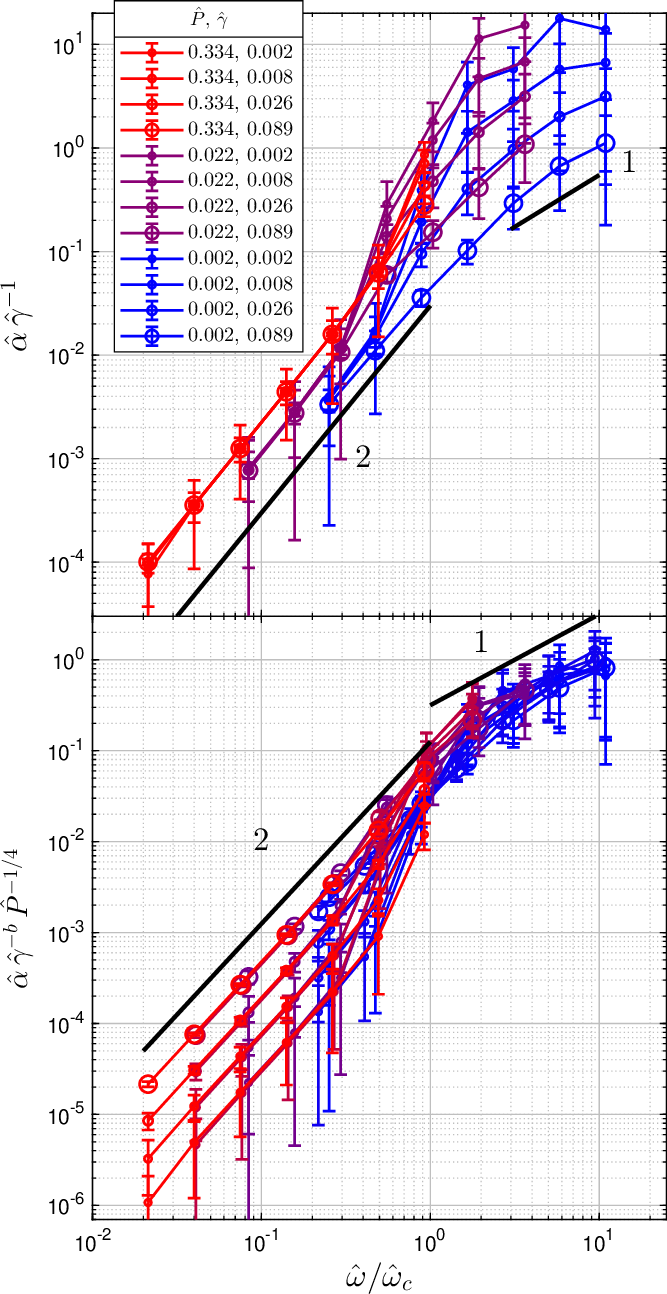}\\[6pt]
    \caption{
    Compression wave attenuation normalized by $\hatg$ (top) and $\hat\gamma^{b}\hatp^{1/4}$ (bottom) versus $\hatw/\wc$, for selected values of $\hatp$ and $\hatg$ (see legend; color indicates $\hatp$, red to blue from high to low).
    Below $\wc$, $\hata/\hatg$ collapses curves of the same $\hatp$ onto a single line with slope 2, confirming $\hata \propto \hatg\hatw^2$, but a residual pressure and sublinear damping dependence remains, in contrast to the full collapse observed for shear waves (Fig.~\ref{fig:shearAttenuation}) in this frequency region.
    Above $\wc$, the $\hata/\hatg$ curves spread, indicating that the linear $\hatg$-dependence breaks down.
    (Bottom) We rescale an effective weak sublinear power law $\hat\gamma^{b}\hatp^{1/4}$, similar to Fig.~\ref{fig:shearAttenuation} for $b<a$. 
    This collapses curves of different $\hatp$ at fixed $\hatg$ below $\wc$, yielding $\hata \propto \hatg^b\hatw^2$.
    We note this $\hat{P}^{1/4}$ scaling is similar to the wavespeed-pressure scaling in Fig.~\ref{fig:wavespeedCombined}, but the connection is unclear. 
    }
    \label{fig:compAttenuation}
\end{figure}

Figure~\ref{fig:compAttenuation} shows $\hata(\hatw)$ for compression waves, normalized by $\hatg$ (top) and $\hat\gamma^b\hatp^{1/4}$ where $b<1$ (bottom), for a range of $\hatp$ and $\hatg$.
Below $\wc$, the data recover the same linear $\hatg$-dependence seen for shear waves, but a residual pressure dependence remains.
The bottom panel rescales by $\hat\gamma^{b}\hatp^{1/4}$ with $b<1$, collapsing curves of different $\hatp$ below $\wc$ and yielding $\hata \propto \hatg^b\hatw^2\hatp^{1/4}$.
Above $\wc$, the $\hatg$-dependence drops to an effective sublinear power law, leaving $\hata \propto\hat\gamma^{b} \hatw\hatp^{1/4}$, weakly dependent of grain-contact damping.
Although the scattering nature of the modes above $\wc$ makes $\hata$ increasingly difficult to resolve at low $\hatg$, the data have been passed through the quality filtering described in Appendix~\ref{app:dft}, and the collapse across multiple values of $\hatp$ and $\hatg$ provides confidence in the reported exponents.
The strongly reduced $\hatg$-dependence above $\wc$ for compression waves, with $b\lessapprox a<1$, is distinct from the sublinear $\hata \propto \hatg^{a}$ behavior observed for shear waves
As with shear waves, the $\hatg$-dependence in the scattering regime is not a simple power law: the curves above $\wc$ in Fig.~\ref{fig:gammaAtten}(a) change shape with $\hatg$ in addition to shifting.
The exponent $b$ therefore captures the overall sublinear trend rather than an exact scaling.

These shear and compression results confirm the central claim of this work: the damped vibrational modes of the packing govern acoustic attenuation, and the threshold $\wc = \hatp^{1/2}$, set entirely by proximity to jamming, marks the boundary between viscous-like dissipation and a scattering-dominated regime that loses sensitivity to grain-contact damping.
The fact that this transition appears in both wave types, with the same threshold and the same change in scaling, points to the disordered mode structure of the packing as the controlling mechanism.
This mechanism provides an explanation for the $\hata \propto \hatw$ scaling observed in sediment acoustics~\cite{zhou_low-frequency_2009,SAX99_overview,Buckingham1997}, one that emerges directly from the granular structure rather than from fitted continuum models.

\section{Discussion \label{sec:Discussion}}

Motivated by the acoustic properties of submerged granular media, we have used discrete-element simulations to study acoustic attenuation and wavespeed for both shear and compression waves in dissipative, jammed granular packings. We varied pressure $\hatp$, frequency $\hatw$, and grain-contact damping $\hatg$.
A central method of this work is the analysis of damped vibrational modes in the attenuation–frequency plane $(\hat \beta_i,\hat\omega_i)$, which differs from the more common approach of using the density of states. This allows us to compare with the measured $\hata(\hat\omega)$ for propagating waves in our numerical simulations.
The damped vibrational modes show two regimes separated by $\wc$: plane-wave-like modes with $\hat\beta_i \propto \hatg\hatw_i^2$ below $\wc$, and scattering modes with weakly frequency-dependent $\hat\beta_i$ above $\wc$.
Wavespeed is largely insensitive to $\hatg$ and recovers the known $\hatp^{1/2}$ scaling of the elastic moduli in both wave types.
Below $\wc$, attenuation is viscous-like: $\hata \propto \hatg\hatw^2$ for both wave types, with coherent oscillations and $A_\perp \ll A_\parallel$, as shown in Figs.~\ref{fig:compAmpPhase} and~\ref{fig:ampPhaseShear}.
Above $\wc$, the excess low-frequency modes produce anomalously high attenuation largely insensitive to grain-contact damping, with $\hata \propto \hat \gamma^b\hatw \hat P^{1/4}$ for compression waves and $\hata \propto \hatg^{a}\hat \omega$ with $a<1$ for shear waves.
The key of the JNS framework is that the pressure-dependent threshold $\wc = \hatp^{1/2}$ separates two fundamentally distinct attenuation regimes.
The results presented here are restricted to the regime $\hatg \ll 1$; when $\hatg \gtrsim 0.1$ grain-contact damping becomes strong enough to modify the overall mode structure, as depicted in rightmost panels in Fig.~\ref{fig:eigen_all}, and the scaling arguments developed here are expected to break down. This regime is likely more relevant to, e.g., foams or emulsions, and we leave it for future study.

These results have direct relevance to the long-standing problem of $\alpha \propto \omega$ scaling in sediment acoustics.
Figure~\ref{fig:experimentalData} shows measured spatial attenuation coefficients from multiple field and laboratory experiments in marine sediments, spanning frequencies from tens of Hz to 1 MHz.
At low frequencies (below roughly 1 kHz), the data follow $\alpha \propto f^2$, consistent with viscous dissipation.
At higher frequencies, the data transition to a nearly linear scaling.
This crossover, and the nearly linear high-frequency scaling, is the central question that motivates this work.
The Biot-Stoll and related viscous models do not produce linear $\omega$ attenuation from first principles, and the Grain Shear theory requires ad hoc 
non-linear contact laws to do so. 
The JNS framework requires no such assumptions: linear attenuation emerges naturally from the disordered contact network with standard linear dashpot dissipation.
Our results suggest a different and more physically based approach: the disordered mode structure of a jammed packing near the jamming transition naturally produces $\hata \propto \hatw$ in the scattering regime, even with linear forces at grain-grain contacts.
The threshold $\wc = \hatp^{1/2}$ provides a predictor for the crossover frequency, expressed directly in terms of the confining pressure and grain stiffness, which can also be interpreted as the stiffness of the particles.

Although our results use packings formed by 2D frictionless disks, we can estimate whether the crossover frequency $\wc = \hatp^{1/2}$ is consistent with the data in Fig.~\ref{fig:experimentalData}, where the crossover frequency appears to be between $1$ and $5$~kHz.
From the definition in Table ~\ref{tab:dimensionless_quantities}, we note that $\hatp$ can also be interpreted as the ratio of the particle compression magnitude $\delta_0$ to the particle diameter $d$. The Hertzian force magnitude for compression $\delta_0$ is nonlinear, given by $(4/3) E^* R^{1/2} \delta_0^{3/2}$, where $R=d/2$ is the particle radius and $E^* = E / [2(1-\nu^2)]$ is the effective Young's modulus for material with Young's modulus $E$ and Poisson's ratio $\nu$. Equating this force law with $\sigma_0d^2$, where $\sigma_0 \approx \Delta\rho g h$ is the characteristic overburden pressure at depth $h$ and $\Delta\rho = \rho_{\text{grain}} - \rho_{\text{water}}$ is the buoyant mass density, we obtain $\hatp\approx \delta_0/d = \left[ 3(1-\nu^2)\sigma_0 / (4E) \right]^{2/3}$.
The critical frequency is then $f_c = \wc \omega_0 /2\pi$, where $\omega_0= \sqrt{\kappa/m}$.
The spring constant $\kappa$ represents the combination of the grain and fluid stiffness, which can be estimated from the measured macroscopic wave speed $c$, such that $\kappa = m c^2/d^2$, where $m$ is the effective grain mass (including the surrounding liquid).
The fact that $c$ is very close to the wave speed in pure water suggests that the fluid stiffness dominates.
This yields a characteristic frequency $\omega_0 = \sqrt{\kappa/m} = c/d$.
The dimensional crossover frequency is therefore 
\begin{equation}
    f_c \approx \frac{c}{2\pi d} \left[ \frac{3}{4}(1-\nu^2) \frac{\Delta\rho \, g \, h}{E} \right]^{1/3}.
\label{eqn:experimentalFreqCutoff}
\end{equation}
Using $h=2$~cm and typical physical values for water-saturated silica sand ($d = 200\ \mu\text{m}$, $\Delta\rho = 1650\ \text{kg/m}^3$, $E = 70\ \text{GPa}$, $\nu = 0.17$, and $c = 1700\ \text{m/s}$), we find $f_c \approx 2.03$~kHz, which is consistent with the where the data in Fig.~\ref{fig:experimentalData} changes from slope 2 to slope 1. We also note that $f_c$ is weakly dependent on $h$. We also note that the Hefner and Williams~\cite{hefner2006sound} data shown in Fig.~\ref{fig:experimentalData} uses both water and silicone oil, which has a viscosity of roughly 100 times that of water. The resulting tenfold increase in $\alpha$ is also consistent with the sublinear scaling we observe for $\alpha$ versus $\gamma$ in the scattering regime.

This jamming-based origin for $\alpha\propto f$ has broad implications across physics and engineering disciplines.
First, the anomalous excess modes (the boson peak), which have been studied with $\hatg = 0$, persist when $\hatg>0$ damping and dominate dissipation when $\hatw>\wc$.
Our results suggest that ad hoc, non-Newtonian dissipation laws are not required to explain linear attenuation.
Instead, it emerges naturally from the geometry of the disordered contact network.
In the geophysical and acoustic engineering fields, the JNS framework may provide a useful bridge between grain-scale simulations and phenomenological models, enabling more physics-based estimates grounded in the behavior of damped, jammed packings.

The simulations presented here use networks generated via 2D frictionless packings, whereas marine sediments are obviously three-dimensional and polydisperse with non-negligible friction between grains.
However, we expect the qualitative features of the JNS framework to extend to three dimensions, although confirming this remains a target for future work.
The scaling exponents governing linear response are known to be the same in 2D and 3D~\cite{ohern2003jamming,van2009jamming}.
Differences between 2D and 3D packings arise from nonlinear or geometric effects, such as steady-state flow behavior, which do not affect the linear wave propagation and the crossover frequency $\wc$ studied here.
However, friction changes the isostatic threshold and can affect the excess low-frequency modes that drive the scattering regime~\cite{somfai2007critical}, potentially shifting the location and sharpness of the $\wc$ crossover. 

Additionally, we have explicitly used only linear forces throughout this paper. Nonlinear contact and dissipation laws, such as those used in Grain Shear model, may be relevant in other settings and are worth exploring in future work.
However, the present results demonstrate that the key features of the data in Fig.~\ref{fig:experimentalData} can be explained from collective effects of a disordered contact network with linear forces.

While we have characterized the modal structure of damped packings across a range of $\hatg$ and $\hatp$, a complete theoretical understanding of the scattering regime remains open.
In particular, the residual pressure dependence of compression waves in both the viscous damping regime ($\hat{\alpha} \propto \hat{\gamma}\hat{\omega}^2 \hat{P}^{1/4}$) and scattering-dominated region ($\hat{\alpha} \propto \hatg^b\hat{\omega} \hat{P}^{1/4}$), as shown in Fig.~\ref{fig:compAttenuation}, and the transition to overdamped modes at large $\hatg$ all warrant further investigation.
The residual $\hatp^{1/4}$ scaling dependence in the compressional attenuation mirrors the pressure scaling of the compressional and shear wavespeed, possibly suggesting a connection between spatial compressional attenuation and wavespeed similar to $\hat \beta = \hat c \, \hat \alpha$ as described in Section \ref{sec:Introduction}.
However, this link is not yet clear and its absence in the shear attenuation data is an open question.

\begin{acknowledgments}
We thank the high-performance computing teams at both UvA/NFWI and NPS. Computations on wave propagation were performed using the University of Amsterdam - Science Faculty (UvA/FNWI) High Performance Computing Facility, a centrally managed computational resource available to UvA/FNWI researchers including faculty, staff, students, and collaborators. Mode computations were performed at NPS on the Hamming supercomputer.
\end{acknowledgments}

\appendix

\section{On-the-Fly Discrete Fourier Transform Algorithm\label{app:dft}}

The wave propagation simulations described in Sec.~\ref{sec:Methods} require extracting the steady-state oscillation amplitude $A_i$ and phase $\phi_i$ at the driving frequency $\omega_D$ for each of the $N$ particles in the channel.
Storing the full displacement trajectory would cost $\mathcal{O}(N \times N_t)$ memory per spatial component, which restricts access to larger system sizes with limited computational resources.
Instead, we accumulate the inner product of each particle's displacement with the complex exponential $e^{-i\omega_D t}$ incrementally during the time-stepping loop, projecting onto exactly $\omega_D$ rather than onto a discrete frequency grid.
This reduces peak memory from $\mathcal{O}(N_t)$ to $\mathcal{O}(1)$ per particle.
Results are benchmarked by full-spectrum DFT calculations on representative simulations.
In all cases examined, the power in the bin at $\omega_D$ exceeds the power in all other bins by at least two orders of magnitude, supporting that energy transfer to other frequencies is negligible and that the attenuation of the single-frequency coefficient fully characterizes the steady-state response.

For a signal $f[n]$ sampled at $N_t$ points with timestep $\Delta t$, the DFT coefficient at frequency index $m$ is

\begin{equation}
    \tilde{F}_m = \frac{1}{N_t}\sum_{n=0}^{N_t-1} f[n]\, e^{-i\, m n \frac{2\pi}{N_t}},
\end{equation}

where the complex amplitude encodes both the oscillation amplitude $|\tilde{F}_m|$ and phase offset $\arg(\tilde{F}_m)$.
Because the positive and negative frequency components $\pm m$ contribute equally to the same physical frequency, the single-sided amplitude is $A = 2|\tilde{F}_m|$.

Since $\omega_D$ is fixed and known exactly throughout each simulation, the projection reduces to accumulating a single inner product per particle with twiddle factor $e^{-i\omega_D t_n}$, where $t_n = n\,\Delta t$.
Unlike a standard DFT whose frequencies are constrained to a discrete grid with bin width $\Delta\omega = 2\pi/T$, this approach projects onto $\omega_D$ exactly.
For our simulations, $f[n] = \Delta u_i^{(d)}[n] = x_i^{(d)}[n] - x_{0,i}^{(d)}$  is the displacement of particle $i$ in direction $d \in \{x,y\}$, the required coefficient is the inner product
\begin{equation}
    C_i^{(d)} = \sum_{n=0}^{N_t-1} \Delta u_i^{(d)}[n]\, e^{-i\omega_D t_n}.
\end{equation}

This is accumulated recursively at each timestep as

\begin{equation}
C_i^{(d)}[n] = C_i^{(d)}[n-1]+ Q[n]\,\Delta u_i^{(d)}[n]\,e^{-i\omega_D t_n}
\end{equation}

where the coefficient is initialized as $C_i^{(d)}[0] = 0$ and  $Q[n] \in \{0,1\}$ is a gating function described in the following paragraph.
A sample counter $N_i$ is incremented alongside the accumulator, and post-simulation normalization yields $\tilde{u}_i^{(d)} = C_i^{(d)}[N_t]/N_i$.

The gating function $Q[n]$ is designed to ensure that only steady-state oscillation is accumulated, excluding both the pre-arrival transient and the high-amplitude soliton launched at $t=0$ when the wall begins oscillating.
For bulk particles, the accumulator opens only after two simultaneous conditions are met: a time gate $n > x_{0,i}/(c_0\,\varepsilon_{\rm arr}\,\Delta t)$ with $\varepsilon_{\rm arr} = 1.2$, which blocks the faster-traveling soliton; and a motion gate $|\Delta u_i^{(x)}[n]| > \varepsilon_{\rm trig} = 10^{-6}$, which confirms physical arrival of the driven wave.
Once both conditions are satisfied, accumulation is further delayed by $N_{\rm delay} = 3$ complete driving periods to allow the wavefront transient to decay.
Driven wall particles are treated separately: their accumulators are active from $n=1$ and span the full simulation duration.

To minimize spectral leakage, each particle's window is constrained to span exactly $n_{\rm cyc}$ complete driving periods after its onset time $n_{\rm on}$,

\begin{equation}
    n_{\rm cyc} = \left\lfloor \frac{N_t - n_{\rm on}}{N_{\rm period}} \right\rfloor,
    \qquad
    N_{\rm period} = \mathrm{round}\!\left(\frac{2\pi}{\omega_D\,\Delta t}\right).
\end{equation}

Particles accumulating fewer than $N_{\rm min} = 5$ complete cycles are excluded from the attenuation and wavenumber fits described in Sec.~\ref{sec:Methods}.

The steady-state amplitude and phase are $A_i = 2|\tilde{u}_i^{(d)}|$ and $\phi_i = \arg(\tilde{u}_i^{(d)})$, consistent with the model displacement field in Sec.~\ref{sec:Methods}.
Spatial attenuation $\alpha$ and wavenumber $k$ are extracted by linear least-squares fits to $\ln A_i$ and $\phi_i$ versus equilibrium position $x_{0,i}$, and the phase velocity follows as $c_\phi = \omega_D/k$.

\section{Data Aggregation and Quality Filtering\label{app:data_filtering}
}
After all simulations are complete, individual output files are combined into a single dataset.
The data are then passed through three sequential quality filters before analysis.

Particles whose oscillation amplitude falls below a threshold $A_{\min} = 3 \times 10^{-7}$ are removed from each run's amplitude and phase vectors.
This excludes particles that have not been meaningfully excited --- either because the wave has not reached them or because their signal has decayed below the noise floor.
After truncation, the attenuation coefficient $\alpha$ and wavenumber $k$ are re-fitted by linear least squares to the surviving data, and the $R^2$ is recorded for both fits.

To exclude the far-field region where the wave has attenuated below reliable measurement, each run's data are further truncated to the first fraction $r$ of the spatial range,
\begin{equation}
    x_{\text{cutoff}} = x_{\min} + r\,(x_{\max} - x_{\min}),
    \qquad r = 0.5,
\end{equation}
with attenuation, wavenumber, and $R^2$ re-computed from the retained data.

Finally, wave speeds are recalculated directly from the phase data for both the $x$- and $y$-displacement components.
For each run and each direction, the phase vector is re-sorted by particle distance, re-wrapped to $[0, 2\pi)$, and spatially unwrapped.
A linear fit $\phi = k\,x + \phi_0$ is applied to the cleaned phase data, and the phase velocity is updated as $c_\phi = \omega_D / k$.
This step ensures that wave speeds are consistent with the filtered amplitude and phase data produced by the preceding cleaning steps.

\bibliography{references}

\end{document}